\documentclass[aps,pre,amsmath,amssymb,floatfix,nofootinbib,preprint]{revtex4-1}
\usepackage{graphicx}
\usepackage{amsmath}
\usepackage{multirow}
\usepackage{chemarr}
\begin{document}
\title{Quantitative Analysis of a Transient Dynamics of a Gene Regulatory Network}
\author{JaeJun Lee and Julian Lee}
\email{jul@ssu.ac.kr}
\affiliation{Department of Bioinformatics and Life Science, Soongsil University, Seoul, Korea}
\date{\today}
\begin{abstract}
In a stochastic process, noise often modifies the picture offered by the mean field dynamics. In particular, when there is an absorbing state, the noise erases a stable fixed point of the mean field equation from the stationary distribution, and turns it into a transient peak. We make a quantitative analysis of this effect for a simple genetic regulatory network with positive feedback, where the proteins become extinct in the presence of stochastic noise, contrary to the prediction of the deterministic rate equation that the protein number converges to a non-zero value. We show that the transient peak appears near the stable fixed point of the rate equation, and the extinction time diverges exponentially as the stochastic noise approaches zero. We also show how the baseline production from the inactive gene ameliorates the effect of the stochastic noise, and interpret the opposite effects of the noise and the baseline production in terms of the position shift of the unstable fixed point. The order of magnitude estimates using biological parameters suggest that for a real gene regulatory network, the stochastic noise is sufficiently small so that not only is the extinction time much larger than biologically relevant time-scales, but also the effect of the baseline production dominates over that of the stochastic noise, leading to the protection from the catastrophic rare event of protein extinction. 
\end{abstract}
%\pacs{05.70.Fh, 05.70.Ln, 02.50.Ey, 64.60.an}
\maketitle
%\newpage
\section{Introduction}
The probability of a rare event in stochastic reaction processes has been a subject of much interest and extensive studies~\cite{assaf,meerson1,assaf4,mobilia5,assaf6,
assaf9,lohmar10,gott13,assaf21,mendez25,smith26,assaf27,assaf28,assaf29,assaf33,assaf34,
doering38,khasin7,black11,khasin16,gabel17,
kamen31,assaf32,meerson13,munoz,munoz2,
kamenev}. Such an event can drastically modify the picture provided by the mean field dynamics, if that event brings the process to an absorbing state. A representative example is the extinction of a population or a disease~\cite{assaf,meerson1,assaf4,mobilia5,
assaf6,assaf9,lohmar10,gott13,assaf21,mendez25,
smith26,assaf27,assaf28,assaf29,assaf33,assaf34,
doering38,khasin7,black11,khasin16,gabel17,
kamen31,assaf32,meerson13}. Under the assumption of an isolated population, the state of the vanishing population or disease is an absorbing state. In this case, even when the mean-field dynamics predicts that the whole or infected population reaches a non-vanishing stationary value, called the stable fixed point, a finite probability flux into the absorbing state leads to an eventual extinction of the population or disease. The stable fixed point is converted into a transient state by the stochastic noise in this case, implying the mean-field description is valid during a finite time, and eventually breaks down.  
 Obviously, the extinction event can be prevented by removing the absorbing state. This can be done by introducing the influx of population or disease, so that the state of the vanishing population or disease is not an absorbing state any more. The rate of such an influx determines the relative dominance of the the mean-field stable fixed point versus the state of vanishing population or disease on the stationary distribution~\cite{meerson13}.

A gene regulatory network with positive feedback~\cite{jong}, shares the same qualitative features as the models of population dynamics or epidemics discussed above, in that the protein activates its own production by binding to the DNA: when there is no protein production from the inactive gene, the state of the vanishing number of proteins becomes an absorbing state. A small amount of the protein production from the inactive gene, called the baseline production, plays the role of population (disease) influx in the case of population dynamics (epidemics), in that it removes the absorbing state. Therefore, it is clear that a similar quantitative analysis can be conducted on the gene regulatory network as in the case of the population dynamics or epidemics. Although the role of stochastic noise in gene regulation has been a focus of much interest recently~\cite{elo,harley,kepler,walczak,fried,loinger,
steve1,grima,vandecan,aquino,kumar,hao2,taba1,
taba,och,taba3,paj,sung1,sung,new1,new2,new3,new4,new5,new6,new7,new8,mac,assaf12,bress8,
bress17,new14,earn18,assaf19,vardi20,robert22,
new24,lipshtat,Artyomov,elf}, it is difficult to find a quantitative analysis of how the stochastic noise turns a stable fixed point
 into a transient state, and how the baseline production rescues the proteins from being extinct, in a gene regulatory network.

In this work, we will consider the simplest form of genetic regulatory network with positive feedback and obtain the time-dependent distribution as a numerical solution of the chemical master equation. We also obtain an analytic solution under the assumption of appropriate time-scale separations. We indeed see that the stable fixed point of the deterministic mean-field dynamics turns into a transient peak of the probability distribution, and gets erased from the stationary distribution in the absence of the baseline production.  We then compute the time-scale for the leakage of the probability to the absorbing state, and find that the leakage time increases as the stochastic noise decreases, making the deterministic equation valid for longer time duration. We then analyze how the baseline production from the inactive gene ameliorates the effect of the stochastic noise by removing the absorbing state. We show that the opposite effects of the stochastic noise and the baseline production can be explained in terms of the position shift of the unstable fixed point.
 
The order of magnitude estimates using biological parameters suggest that for a real gene regulatory network, the stochastic noise is sufficiently small so that not only is the leakage time much larger than biologically relevant time-scales, but also the effect of the baseline production dominates over that of the stochastic noise, leading to the protection from the catastrophic rare event of protein extinction.
\section{The model}
The model we consider is the simplest genetic regulatory network with  positive feedback loop. We consider a protein $X$ that binds to the DNA to activate its own production:
\begin{eqnarray}
D + X &\xrightleftharpoons[k_1]{k_0}& D^*\nonumber\\
D^* &\xrightarrow{a}& D^* + X\nonumber\\
D &\xrightarrow{a \epsilon}& D + X\nonumber\\
X &\xrightarrow{b}&  \varnothing\nonumber\\
D^* &\xrightarrow{b \rho}&  D
 \label{auto}
\end{eqnarray}
where $D^*$ and $D$ denote the DNA with protein bound and unbound, respectively. Although $X$ is produced from $D^*$ or $D$ via transcription and translation, we assume that they can be approximated as a one-step process. We assume that the degradation rate of the bound protein is not greater than that of the free protein, so that $0 \le \rho \le 1$. Although most of the results presented are for $\rho=0$, the value of $\rho$ does not affect the qualitative feature of the results. The nonnegative number $\epsilon$ parametrizes the  rate of transcription from the inactive gene, called the baseline production~\footnote{This is also called the transcriptional leakage, but we will refrain from using this terminology because we will be  using the word leakage in quite the opposite sense.}~\cite{fried,taba,paj,weber,minaba,guo}. Because we are considering the case of a positive feedback, the value of $\epsilon$ is restricted to be $0 \le \epsilon < 1$. 
 
\section{The deterministic rate equation}
We assume that the time-scale of equilibration between $D$ and $D^*$ is much shorter than other relevant time scales, so that they can be assumed to be equilibrated instantly. We also assume that the number of $X$ molecules is large enough so that its fluctuation can be neglected. Then the probability that the DNA is bound to a protein molecule is given by
\begin{eqnarray}
p(D^*) &=& \frac{k_0 x}{k_0 x + {\tilde k}_1} = \frac{x}{x + \tilde r} \nonumber\\
 p(D) &=& 1-p(D^*)   = \frac{\tilde r}{x + \tilde r}
\end{eqnarray}
at any instant of time, 
where $\tilde r \equiv {\tilde k}_1/ k_0$ with $k_0$ and ${\tilde k}_1$ being the binding and unbinding rates between the protein $X$ and the DNA, and $x$ is the concentration of the protein $X$~\footnote{The concentration is defined as $x \equiv m/{\bar m}$, where $m$ is the number of proteins and $\bar m$ is a large number chosen to be of the order of average number of proteins~\cite{kepler}. The rates $\tilde r$, $\tilde k_1$, and $\tilde a$ of the rate equation, and the corresponding quantities $r$, $k$, and $a$ in the master equation, are related by $\tilde r = r/\bar m$, ${\tilde k}_1 = k_1/\bar m$, and $\tilde a = a/\bar m$. See appendix \ref{FP} for details.}. The rates for the production and the degradation of $X$ are proportional to $p(D^*) + \epsilon p(D)$ and $x$, respectively, leading to the deterministic rate equation
\begin{equation}
\dot x = \frac{\tilde a (x + \tilde r \epsilon)}{x+\tilde r} - b x, \label{de2}
\end{equation}
describing the mean-field dynamics of $x$ where its fluctuation is neglected. The effect of the degradation of the bound protein is negligible in this limit, as shown in Appendix ~\ref{FP}, and therefore $\rho$ does not appear in Eq.(\ref{de2}).

Although physically $x \ge 0$, we first obtain the fixed points of Eq.(\ref{de2}) and examine their stability without such restriction for the convenience of analysis.  
Fixed points of Eq.(\ref{de2}) are obtained by setting $\dot x$ to zero, which is equivalent to solving the equation 
\begin{equation}
x^2+(\tilde r  - \tilde a/b) x -r \tilde a \epsilon/b = 0. \label{fe1}
\end{equation}
The roots of Eq.(\ref{fe1}) are
\begin{equation}
x_\pm = \frac{1}{2} \left[ \frac{\tilde a}{b} - \tilde r \pm \sqrt{(\frac{\tilde a}{b} - \tilde r)^2 + \frac{4 \tilde a \tilde r \epsilon}{b} }\right], \label{trt}
\end{equation}
whose stability can be analyzed by expanding Eq.(\ref{de2}) up to linear order in $\delta x \equiv x - x_\pm$:
\begin{equation}
\dot {\delta x} = H(x_\pm) \delta x \equiv \left[ -b x + \frac{\tilde a (x + \tilde r \epsilon)}{x+\tilde r} \right]'_{x_\pm} \delta x = \left( -b  + \frac{\tilde a \tilde r (1-\epsilon)}{(x_\pm+\tilde r)^2} \right) \delta x .  
\end{equation}
Because $x_+ + \tilde r \ge x_-  +\tilde r$ and $(x_+ + \tilde r)(x_- + \tilde r) =   \tilde a \tilde r (1-\epsilon)/b$,
we get 
\begin{equation}
 (x_+ +  \tilde r)^2 \ge  \tilde a \tilde r (1-\epsilon)/b \ge    (x_- + \tilde r)^2,
\end{equation}
from which we get $\tilde a \tilde r (1-\epsilon)/(x_+ +  \tilde r)^2 \le b$ and $\tilde a \tilde r (1-\epsilon)/(x_- +  \tilde r)^2 \ge b$ where the relations are satisfied as equalities only when $\epsilon=0$ and $\tilde a/b = \tilde r$ so that $x_+ = x_-$. Therefore, $H(x_+)<0$ and $H(x_-)>0$ if $\tilde a/b \ne \tilde r$ or $\epsilon>0$, whereas $H(x_\pm=0) = 0$ for $\epsilon = \tilde a/b - \tilde r = 0$. 
Consequently, we see that $x_+$ and $x_-$ are stable and unstable fixed points respectively, for the former case. For the latter case, the stability of $x_\pm=0$ is analyzed by expanding Eq.(\ref{de2}) up to second order in $\delta x$, where we find that 
\begin{equation}
\dot {\delta x} = H'(0) \delta x^2 =  - \frac{2 \tilde a \tilde r (1-\epsilon)}{\tilde r^3} \delta x^2 <0 .  
\end{equation}
Therefore, $x=0$ is a half-stable fixed point because $\dot{\delta x}/\delta x<0$ for $\delta x>0$ and $\dot{\delta x}/\delta x>0$ for $\delta x<0$. 

Now we restrict ourselves to the physical region of $x \ge 0$. When $\epsilon >0$, $x_+ > 0$ and $x_-<0$, and therefore only $x_+$ lies in the physical region. Therefore $x_+>0$ is not only the unique stable fixed point, but it is also the unique fixed point. For the case of $\epsilon=0$, we have $x_+>0$ and $x_-=0$ if $\tilde a/b > \tilde r$, and $x_+=0$ and $x_-<0$ if $\tilde a/b < \tilde r$. Therefore, $x_+>0$ and $x_-=0$ are the stable and unstable fixed points if  $\tilde a/b > \tilde r$, whereas $x_-=0$ is the unique stable fixed point that is also the unique fixed point if $\tilde a/b < \tilde r$. Finally, the unique fixed point at $x=0$ for $\epsilon=0$ and $\tilde a/b = \tilde r$ is also a stable fixed point because now we allow only $\delta x$ with positive sign. 

The results are summarized as follows: 

\noindent
{\bf i}) $\epsilon>0$\\
$x_+=\frac{1}{2} \left[ \frac{\tilde a}{b} - \tilde r + \sqrt{(\frac{\tilde a}{b} - \tilde r)^2 + \frac{4 \tilde a \tilde r \epsilon}{b} }\right] > 0$ is the unique fixed point that is stable.

\noindent
{\bf ii}) $\epsilon=0$ and $\tilde a/b > \tilde r$\\
There are two fixed points. $x_+ = \tilde a/b - \tilde r >0$ is the stable fixed point and $x_-=0$ is the unstable fixed point.

\noindent
{\bf iii)} $\epsilon=0$ and $\tilde a/b \le \tilde r$\\ 
$x_+=0$ is the unique fixed point that is stable.

Note that the case for $\epsilon>0$ can be understood in terms of position shift of a fixed point, starting from $\epsilon=0$ cases. Starting from the case (ii), turning on non-zero $\epsilon$ shifts the position of the unstable fixed point $x_-=0$ to an unphysical value of $x_-<0$, leaving only the non-zero stable fixed point $x_+$ in the physical region, leading to the case (i). If we start from the case (iii), the position of the unique stable fixed point $x_+=0$ is shifted to $x_+>0$ by turning on non-zero $\epsilon$, again leading to the case (i). The position shift of the fixed point at $x=0$  by the baseline production and the stochastic noise will be discussed again later (Section ~\ref{pos}), where we will show that their effects are opposite to each other.

 The case of $\epsilon=0$ and $\tilde a/b > \tilde r$ is of much interest, because the features of the stationary distribution  obtained from the stochastic equation is quite the opposite to the picture offered by the deterministic rate equation. Because $x_+$ is the unique stable fixed point of the deterministic rate equation, $x \to x_+$ in the limit of $t \to \infty$, even if the initial value of $x$ was close to $x=0$. This seems to suggest that in the context of the stochastic dynamics, the stationary distribution should have a peak near $x_+$. However, as will be shown next, the stationary distribution is concentrated at $x=0$ which was predicted to be an unstable fixed point, and has vanishing probability at the stable fixed point. We will see that introducing a nonzero value of $\epsilon$ ameliorates this effect, but as long as its value is sufficiently small, the probability distribution is still dominated by $x=0$.

\section{Chemical Master Equation}
\begin{figure}
\includegraphics[width=0.8\textwidth]{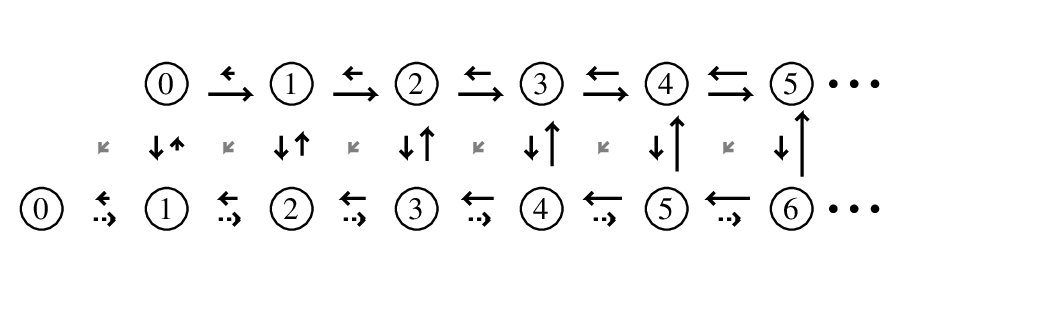}
\caption{An example of the rates and the states of the Markov model corresponding to Eq.(\ref{master1}). The lengths of the arrows are the magnitudes of the transition rates between the states. The numbers in the circle are the numbers of {\it free} protein molecules. The horizontal arrays of states at the top and the bottom are the bound and the free modes, the sets of states with protein-bound and free DNA, respectively. The short gray diagonal arrows exist for $\rho > 0$. The dotted arrows indicate the baseline production, which is absent for $\epsilon=0$.}
\label{rate2} % caption for the whole figure
\end{figure}

There are two sources of stochastic noise: The fluctuation of the protein numbers, and the fluctuation between the bound and unbound states of the DNA~\footnote{The latter can also be considered as the fluctuation in the molecule numbers, the number of  unbound(bound) DNA molecule fluctuating between 0 and 1. }. To fully incorporate the effects of these fluctuations, we should consider the probability  $P(m,n,t)$ that the number of free and bound protein molecules are $m (=0,1, \cdots)$ and $n (=0,1)$ at time $t$. The chemical master equation describing the time-evolution of $P(m,n,t)$ is
\begin{eqnarray}
\dot P(m,0) &=&  -  k_0 P(m,0) m +   k_1 P(m-1,1)   + \epsilon a P(m-1,0) - \epsilon a P(m,0)   \nonumber\\
&& +  b P(m+1,0)(m+1) -  b P(m,0) m  + \rho b P(m,1)\nonumber\\
\dot P(m,1) &=&  k_0 P(m+1,0)(m+1) -  k_1 P(m,1)     +   a P(m-1,1) - \ a P(m,1) \nonumber\\
&& +  b P(m+1,1)(m+1) -  b P(m,1) m  - \rho b P(m,1). \label{master1}
\end{eqnarray}
where it is to be understood that $P(m,n,t) \equiv 0$ whenever $m<0$. Let us call the states with $n=0$ and $n=1$ as the free and the bound mode, the set of states with free and protein-bound DNA, respectively.  The Markov chain corresponding to Eq.(\ref{master1}) is shown in Figure \ref{rate2}, where we immediately see that $(m,n)=(0,0)$ is an absorbing state for $\epsilon=0$, because once the system enters this state due to stochastic fluctuation, there is no way that it can escape to another state, because no protein molecule can be produced from a free DNA. In other words, 
\begin{equation}
P^{\rm st}(m,n)=\delta_{m,0} \delta_{n,0} \label{deltaftn}
\end{equation}
is not only a stationary solution of Eq.(\ref{master1}) with $\epsilon=0$, but also $P(m,n,t)$ converges to $P^{\rm st}(m,n)$ regardless of the initial condition~\cite{markov}. This is in stark contrast to the picture given by the deterministic rate equation Eq.(\ref{de2}), where the system is predicted to move away from $m=0$ and converges to a state with non-vanishing number of the protein molecules. The situation is different from that of a mutual repressor model where the number of peaks of the stationary distribution differs from that of the stable fixed point only when the stochastic noise is sufficiently large~\cite{lipshtat}. In the current model, the results of the deterministic and the stochastic equations contradict each other for all  parameter values, as long as $\epsilon=0$.

The vanishing probability for states other than $(m,n)=(0,0)$ at $t \to \infty$ comes from the fact that $(m,n)=(0,0)$ is an absorbing state of the system and does not depend on the details of the model~(Appendix \ref{other}). Similar situation has also been encountered in models of population dynamics, epidemics, and low-dimensional percolation, whose stochastic master equations share similar structures as the current model of gene regulatory network, Eq.(\ref{master4})~\cite{assaf,meerson1,assaf4,mobilia5,assaf6,assaf9,lohmar10,gott13,assaf21,mendez25,smith26,
kamenev,assaf27,assaf28,assaf29,assaf33,assaf34,doering38,khasin7,black11,khasin16,gabel17,kamen31,assaf32,munoz,munoz2}. There, it has been found that the stable fixed point becomes a transient peak of a quasi-steady distribution instead of the true stationary one. The same statement can be made for the current model by analyzing the time-dependent behavior of the probability distribution, as shown next.

\subsection{Numerical computation of time-dependent solutions for $\epsilon=0$}
\begin{figure}
\includegraphics[height=0.7\textheight]{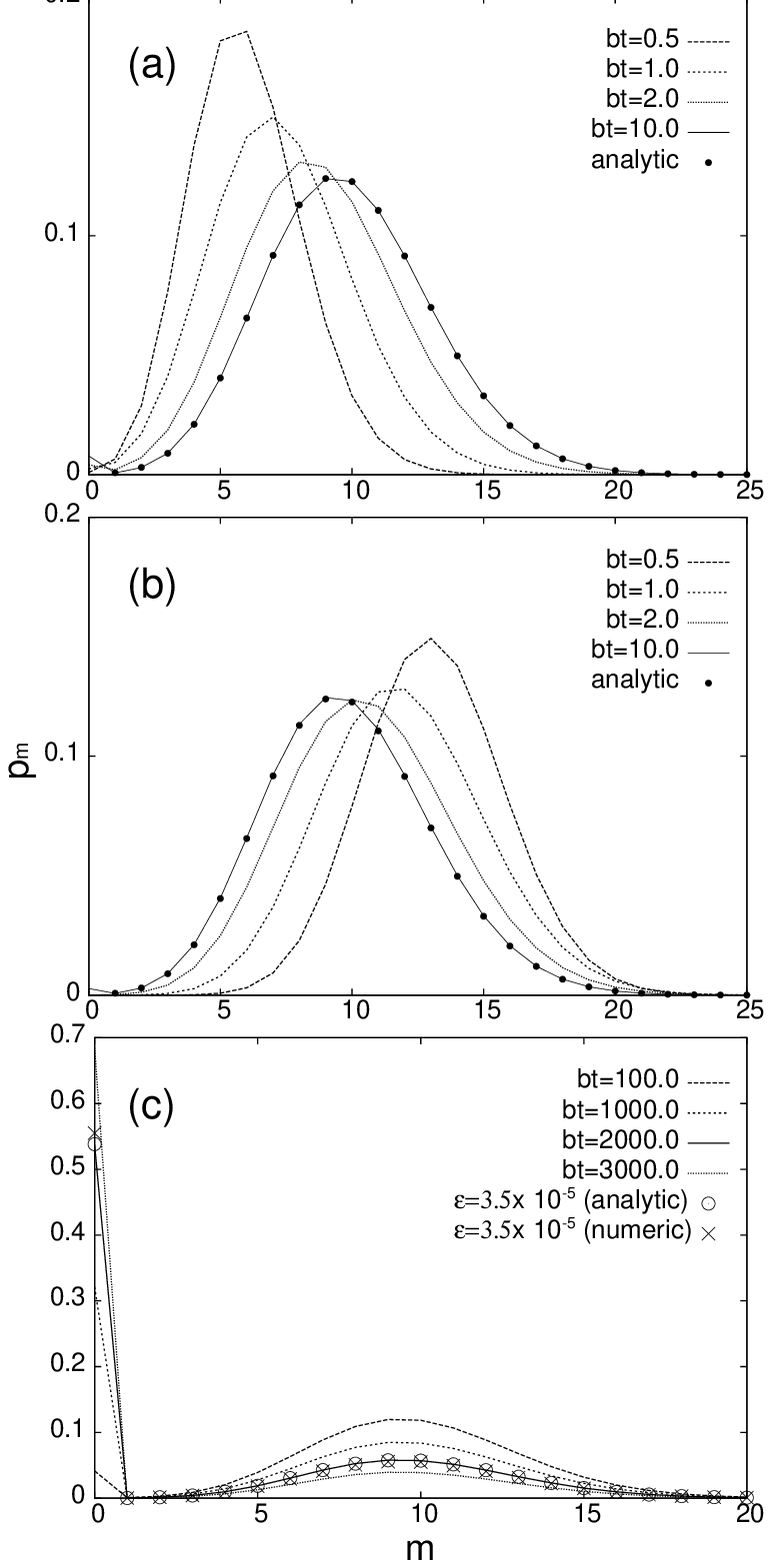}
\caption{The marginal probability distribution $p_m(t)$ plotted as the function of $m$, (a, b) at early times , and (c) at late times, for $a/b=10$,  $k_1 = k_0= 100 b$, $\rho=0$, and $\epsilon=0$. In (a) and (b), 
 $p_m(t)$ at $bt=0.5$, $1.0$, $2.0$, and $10.0$, are drawn. The filled circles are the analytic quasi-steady distribution in Eq.(\ref{qsol}), where the normalization was determined to give the best fit. The initial distributions are given as $P(m,0,0)= 0.2 \delta_{m,4}$ and  $P(m,1,0)=0.8 \delta_{m,3}$ in Figure (a), and $P(m,0,0)= 0.0625 \delta_{m,15}$ and  $P(m,1,0)=0.9375 \delta_{m,14}$ in Figure (b). In Figure (c),  $p_m(t)$ at $bt=100.0$, $1000.0$, $2000.0$, and $3000.0$, are drawn. Stationary distributions for $\epsilon=0.000035$, obtained from the numerical computation and the analytic formula, are also plotted as crosses and circles and, respectively. The other parameters are the same as above, except that $K/b=\infty$ for the analytic solution.     
}
\label{sim} % caption for the whole figure
\end{figure}

Because it is difficult to obtain a general time-dependent solution of Eq.(\ref{master1}) in an analytic form, the equation was solved numerically using the finite-buffer discrete chemical master equation method~\cite{fb1,fb2}, where the state space is truncated to a finite subspace. The state space was truncated so that $m \le 30$, which is a reasonable approximation because $P(30,n, t)<10^{-6}$ at all times, for the initial conditions and the parameters used in the computation. 
%Therefore, the probability distribution obtained from the numerical computation can be essentially considered as the exact result. 
The  master equation was integrated using EXPOKIT package~\cite{EXPOKIT}. For ease of comparison with the results from the next sections, the marginal probability distribution $p_m (t) \equiv P (m,0, t) + P(m-1,1, t)$ was obtained, which is the probability that the total number of proteins, both bound and unbound, is $m$ at time $t$. The   marginal distributions $p_m(t)$  at $t=0.5b^{-1}$, $1.0b^{-1}$, $2.0b^{-1}$, and $10.0b^{-1}$ are shown in Figure \ref{sim} (a) and (b), where the parameters used are $a=10b$, $k_0=k_1=100b$, $\rho=0$, and $\epsilon=0$. The initial conditions are $P(m,0,0)= 0.2 \delta_{m,4}$ and $P(m,1,0)=0.8 \delta_{m,3}$ for Figure \ref{sim}(a) and $P(m,0,0)= 0.0625 \delta_{m,15}$ and $P(m,1,0)=0.9375 \delta_{m,14}$ for Figure \ref{sim}(b). The distribution becomes independent of the initial condition around $t=10 b^{-1}$, and the peak of the distribution for $m>0$ is indeed found at the  stable fixed point of the deterministic rate equation, $m^* = a/b-r = 9$. The peak at the stable fixed point is maintained at later times, but its height decreases due to the leakage of the probability to the state $m=0$, as can be seen in $p_m(t)$ for $t=100.0b^{-1}$, $t=1000.0b^{-1}$, $t=2000.0b^{-1}$, and $t=3000.0b^{-1}$, shown in Figures \ref{sim} (c) for the same parameters. The stable fixed point of the deterministic rate equation has been changed to a transient peak due to the stochastic noise, as expected.

\subsection{Analytic form of the stationary distribution}
\begin{figure}
\includegraphics[width=0.8\textwidth]{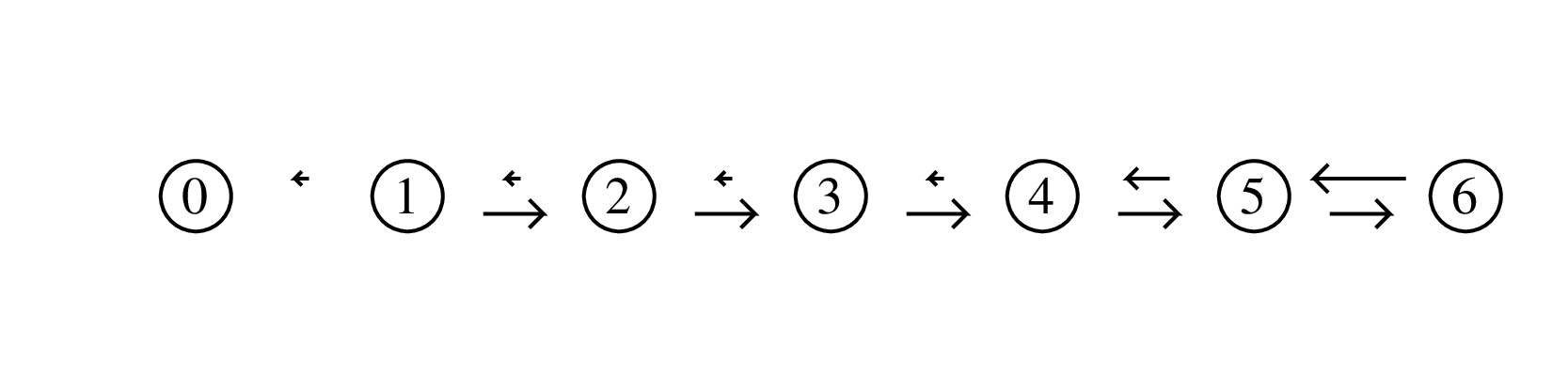}
\caption{An example of the rates and the states of the Markov model corresponding to Eq.(\ref{master4}) for $\epsilon=0$. The lengths of the arrows are the magnitudes of the transition rates between the states. The numbers in the circles indicate the numbers of {\it total} protein molecules, both bound and unbound. In this example, the probability flows to the state $m=5$ on average, but there is also a leakage to $m=0$, whose effect becomes important at late times. }
\label{rate} % caption for the whole figure
\end{figure}

The stationary distribution of the master equation Eq.(\ref{master1}) can be obtained analytically, under the assumption that the rates for the binding and unbinding of the protein molecule to DNA is instantaneous. We first replace the parameters $k_i$ by $K \equiv k_0$ and $r \equiv k_1/k_0$.  Then, in the limit of $K \to \infty$,  we derive the master equation for the marginal probability $p_m (t)$~(Appendix \ref{mdv}):
\begin{eqnarray}
\dot p_m &=& \frac{b (m+1 ) (r + m + \rho )}{m+1  + r} p_{m+1} -  \frac{ b m  (r + m-1 + \rho)}{m  + r} p_{m} -  \frac{ a (m + r \epsilon)  }{m  + r} p_{m}  \nonumber\\
&&+  \frac{a (m-1 +\epsilon r )  }{m-1  + r} p_{m-1}, \label{master4}
\end{eqnarray}
where $p_{-1}(t) \equiv 0$. The corresponding Markov chain is shown in Figure \ref{rate}.

First, we compute the stationary distribution. In general, obtaining an analytic form of the stationary solution is difficult, and one often resorts to additional approximations such as WKB formalism~\cite{assaf,meerson1,assaf4,mobilia5,assaf6,assaf9,lohmar10,gott13,assaf21,mendez25,smith26,
kamenev,assaf27,assaf28,assaf29,assaf33,assaf34,doering38,khasin7,black11,khasin16,gabel17,kamen31,assaf32}. However, the stationary solution of Eq.(\ref{master4}) can be computed exactly, by noting that a stationary distribution of a Markov chain without a cycle must obey a stronger condition called the detailed balance~\cite{kogo,kelly,julpre}, which is
\begin{equation}
\frac{b m (m+r-1+\rho)}{m+r}p^{\rm st}_m = \frac{a (m-1) + \epsilon a r}{m+r-1}p^{\rm st}_{m-1}, \label{db}
\end{equation}
for the Markov chain described by Eq.(\ref{master4}). 
Eq.(\ref{db}) can be solved to obtain the solution:
\begin{equation}
p^{\rm st}_m (\epsilon)= C \left( \frac{a}{b} \right)^{m-1} \frac{(m+r) \Gamma(m+r \epsilon)}{m!  \Gamma(m+r+\rho)}. \label{statsol}
\end{equation}
where $C$ is the normalization constant. When $\epsilon=0$, $\Gamma(m+r \epsilon)$ term in the numerator diverges for $m=0$, and therefore $C=0$ and consequently $p^{\rm st}_m (\epsilon=0) = 0$ for $m>0$, recovering the obvious result: 
\begin{equation}
p^{\rm st}_m (\epsilon=0) = \delta_{m,0}.
\end{equation} 

\subsection{Analytic form of a time-dependent solution for $\epsilon=0$}
Now consider a time-dependent solution of Eq.(\ref{master4}) for $\epsilon=0$. Denoting the transition rate from the state with protein number $m$ to that with $n$ as $k_{m \to n}$, we see that $k_{m \to m+1} = a m / (m+r)$ and $k_{m \to m-1} = b m (r+m-1+\rho)/(m+r)$. Therefore,  $k_{m \to m-1}/k_{m \to m+1} = b(m+r+\rho-1)/a$, and although there is a non-zero probability for transitions in both directions for $m>0$, the most probable direction for transition is the positive direction for $0 < m < a/b+1-r-\rho$, and negative direction for $m > a/b+1-r-\rho$, consistent with the picture provided  by the deterministic rate equation: The particle number converges to a non-zero stable fixed point.  While there is a non-zero probability that the system makes a series of transitions in negative direction to $m=0$ and gets trapped there, the probability for such a rare event can be neglected at early times. Therefore, we assume additional time-scale separation, that $p_0(t)$ is essentially constant during the time-scale where the states with $m>0$ equilibrate among themselves. We have already seen that this assumption is reasonable, by numerically solving the original master equation Eq.(\ref{master1}), but it can also be checked from the analytical solution itself {\it a posteriori}, as will be discussed below.

During the time-scale where the leakage to $m=0$ state is negligible, the dynamics of the states with $m>0$ is described the approximate equation
\begin{eqnarray}
\dot p_m (t) &=& \frac{b (m+1) (r + m + \rho)}{m+1  + r} p_{m+1} (t) -  \frac{ b m  (r + m-1 + \rho)}{m  + r} p_{m} (t) (1-\delta_{m,1}) \nonumber\\
&&-  \frac{ a m  }{m  + r} p_{m} (t)  +  \frac{a (m-1)    }{ m-1  + r} p_{m-1} (t), \label{mrm}
\end{eqnarray}
which is obtained from Eq.(\ref{master4}) with $\epsilon=0$ by blocking the transition from the state $m=1$ to $m=0$. Then the quasi-steady distribution for $m>0$ can be defined as the stationary solution of the modified master equation (\ref{mrm}). The detailed balance condition for Eq.(\ref{mrm}) is again given by Eq.(\ref{db}) with $\epsilon=0$, except that now the value of $m$ is restricted to be positive, leading to the quasi-steady distribution 
\begin{equation}
p^{\rm qs}_m = \tilde C \left( \frac{a}{b} \right)^{m-1} \frac{m+r }{m \Gamma(m+r+ \rho)}. \label{qsol}
\end{equation}
The quasi-steady distribution $p^{\rm qs}_m$ has exactly the same form as the stationary distribution $p^{\rm st}_m$ in Eq.(\ref{statsol}) for $\epsilon=0$, but because $m$ is restricted to be positive values, $\tilde C$ is not zero any more. Once we take the leakage  to $m=0$ state into account, the overall normalization constant $\tilde C$ becomes a slowly decreasing function of time. From the normalization condition $\sum_{m=1}^\infty p^{\rm qs}_m =1 - p_0(t)$, we have
\begin{equation}
p^{\rm qs}_m(t) = (1-p_0(t))\left[\sum_{s=1}^\infty \left( \frac{a}{b} \right)^{s-1} \frac{s+r }{s \Gamma(s+r+ \rho)}\right]^{-1}\left( \frac{a}{b} \right)^{m-1} \frac{m+r }{m \Gamma(m+r+ \rho)} \label{qsol2}
\end{equation}
for $m>0$. 

The local maximum $m^*$ of the quasi-steady distribution is obtained from the condition
\begin{equation}
\frac{p^{\rm qs}_{m}}{p^{\rm qs}_{m-1}} = \frac{a (m-1) (m+r )}{b m (m+r -1) (m+r + \rho -1)} =  1,\label{lomx}
\end{equation}
where it is to be understood that the actual value of $m^*$ should be taken as the integer value close to the real value of $m$ satisfying Eq.(\ref{lomx}).  
In the regime where $m^* \gg 1$, Eq.(\ref{lomx}) reduces to
\begin{equation}
\frac{a }{b (m^*+r)} \simeq  1,
\end{equation}
from which we get 
\begin{equation}
m^* \simeq \frac{a}{b}-r   \gg 1. \label{result1}
\end{equation}
 The deterministic rate equation is written in terms of the concentration $x \equiv m/{\bar m}$, where $\bar m$ is a large number of size $O(m^*)$, and by defining $\tilde a \equiv a/\bar m$ and $\tilde r \equiv r/\bar m$, Eq.(\ref{result1}) is rewritten as
\begin{equation}
x^* \simeq \frac{\tilde a}{b}-\tilde r\quad ({\rm for}\ a/b-r \gg 1). \label{result2}
\end{equation}
Because $m^*\simeq a/b-r$ is the most probable number of protein molecules at early times, it is of the order of average molecules. In this case, the magnitude of the fluctuation is expected to be of order $O(\sqrt{m^*})$, and consequently the relative error is of order $O((m^*)^{-1/2})$~\cite{kampen2,dyken}. Therefore, $(m^*)^{-1/2} \simeq (a/b-r)^{-1/2}$  can be considered as the parameter  
 characterizing the size of the stochastic noise due to the protein number fluctuation, and Eq.(\ref{result2}) tells us that the peak of the quasi-steady distribution is concentrated at the stable fixed point of the deterministic rate equation if the stochastic noise is small.
 
We note that for $m^*=a/b-r \gg 1$, typical values of the transition rates $k_{m \to n}$ between $m>0$ and $n>0$ are much larger than $k_{1 \to 0}$. For $m>0$, we have   $k_{m \to m+1} = a m/(m+r) \sim a m^*/(m^*+r) = a-br \gg   b= k_{1 \to 0} $ by the assumption $a/b-r \gg 1$.  Similarly, $k_{m \to m-1} =  b m (m+r+\rho)/(m+r) \sim  b m^* = b(a/b-r) = a - br \gg   b= k_{1 \to 0}$.  Therefore, states around $m \sim1 $ act as a probabilistic barrier if $a/b - r \gg 1$, and the approximation used in deriving Eq.(\ref{qsol2}) is justified. In fact, the analytic form of the quasi-steady distribution in Eq.(\ref{qsol}) nicely captures the shape of  the actual probability distribution even for $a=10b$, as shown Figures 2(a) and 2(b). In the figures, the quasi-steady solution Eq.(\ref{qsol}) is shown as filled circles, where the overall normalization was adjusted to obtain the best fit with the numerical solution $t=10.0 b^{-1}$. We find excellent agreement regardless of the initial condition. 
 
 We can also obtain the analytic form of $p_0(t)$ that determines the overall normalization $1-p_0(t)$ of the quasi-steady distribution for $m>0$. From Eq.(\ref{master4}), we have
\begin{equation}
\dot p_0(t)=\frac{b (r+\rho)}{r+1} p_1(t). \label{sodum}
\end{equation}
for $\epsilon=0$. Since we are using the quasi-steady state  approximation for $p_m(t)$ with $m>0$, we may substitute $p^{\rm qs}_1(t)$  given in Eq.(\ref{qsol2}) into $p_1(t)$ of Eq.(\ref{sodum}) to get
\begin{eqnarray}
&&\dot p_0(t)  = \frac{b (r+\rho)}{r+1}p^{\rm qs}_1 (t)= \frac{b  }{\Gamma(r+\rho)}\left[\sum_{s=1}^\infty \left( \frac{a}{b} \right)^{s-1} \frac{(s+r )}{s \Gamma(s+r + \rho)}\right]^{-1} (1-p_0(t)),\label{appeq}
\end{eqnarray}
the solution of which is
\begin{equation}
p_0(t) = 1-\exp(-t/\tau_q) \label{expsl}
\end{equation}
where
\begin{equation}
\tau_q^{-1} \equiv \frac{b  }{\Gamma(r+\rho)}\left[\sum_{s=1}^\infty \left( \frac{a}{b} \right)^{s-1} \frac{(s+r )}{s \Gamma(s+r + \rho)}\right]^{-1}.\label{leakrate}
\end{equation}
Eqs.(\ref{qsol2}), (\ref{expsl}) and (\ref{leakrate}) completely specify the analytic form of the time-dependent distribution.

\section{The time-scale separation and the rate of leakage}
 In general, when we construct a matrix ${\bf K}$ whose ($i$,$j$)-th element is given by the transition rate $k_{i \to j}$ of a Markov process, then zero is an eigenvalue of {\bf K},  and all the remaining non-zero eigenvalues are negative~\cite{markov}.  
 Let us denote the negative eigenvalues as $0 > \lambda_1 > \lambda_2 > \cdots $, and call $\lambda_1$ the lowest eigenvalue. These eigenvalues parametrize the multi-exponential convergence of the probability distribution $p_m(t)$ to the stationary one $p^{\rm st}_m$:
\begin{equation}
p_m(t) = p^{\rm st}_m + \sum_{k} v^{(k)}_m \exp(-|\lambda_k| t), \label{muex}
\end{equation}   
where $v^{(k)}_m$ is the $m$-the component of the eigenvector corresponding to the eigenvalue $\lambda_k$, whose normalization is determined by the initial condition. In general, the multi-exponential behavior in Eq.(\ref{muex}) can be approximated as a single exponential:
\begin{equation}
p_m(t) \simeq p^{\rm st}_m +  v^{(1)}_m \exp(-|\lambda_1| t), \label{sexp}
\end{equation}
if $t \gg |\lambda_1|$, in which case $| p^{\rm st}_m - p_m(t) |  \ll 1$. However, if we have a time-scale separation so that $|\lambda_1| \ll |\lambda_k|$ for $k \ge 2$, then the single-exponential  Eq.(\ref{muex}) is a good approximation even for $t \sim |\lambda_1|^{-1}$, where the deviation $p^{\rm st}_m - p_m(t)$ is sizeable.

\begin{figure}
\includegraphics[width=\textwidth]{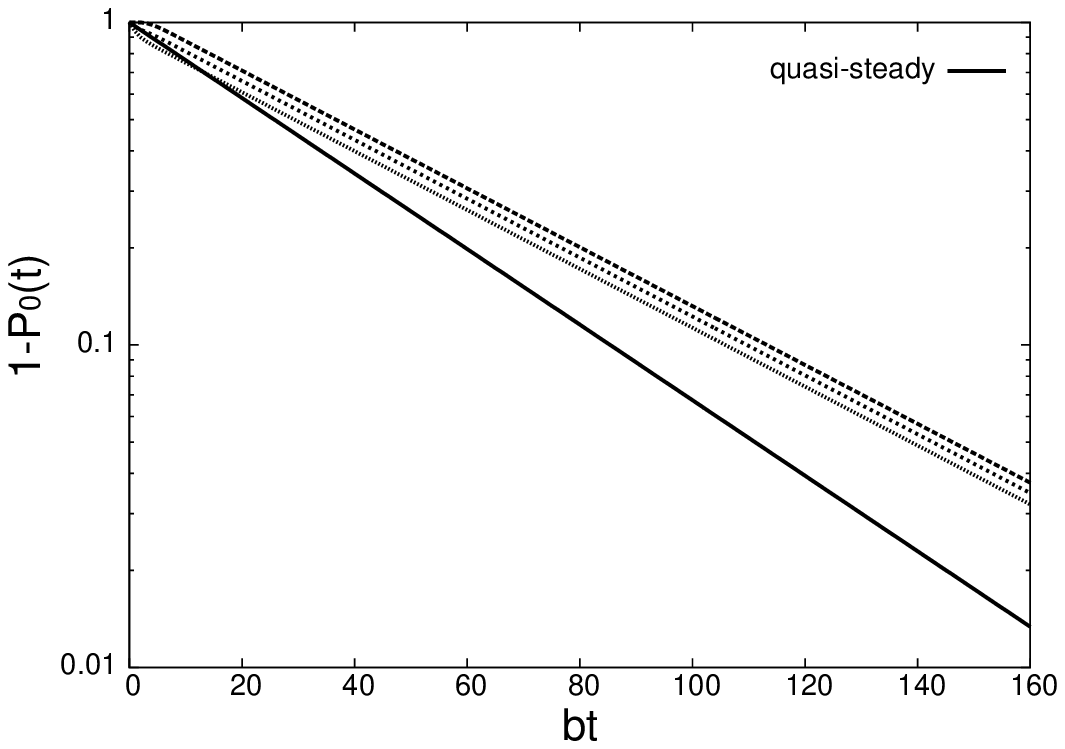}
\caption{The graphs of $1-p_0(t)$ for $a/b=5$, $r=1$, $K/b=\infty$, $\rho=0$, and $\epsilon=0$, where the vertical axis is in log scale. The dashed lines are the results from the numerical computation, with several different initial distribution. The solid line is the result from the analytic expressions in Eqs.(\ref{expsl}) and (\ref{leakrate}).
}
\label{expf5} % caption for the whole figure
\end{figure}

\begin{figure}
\includegraphics[width=\textwidth]{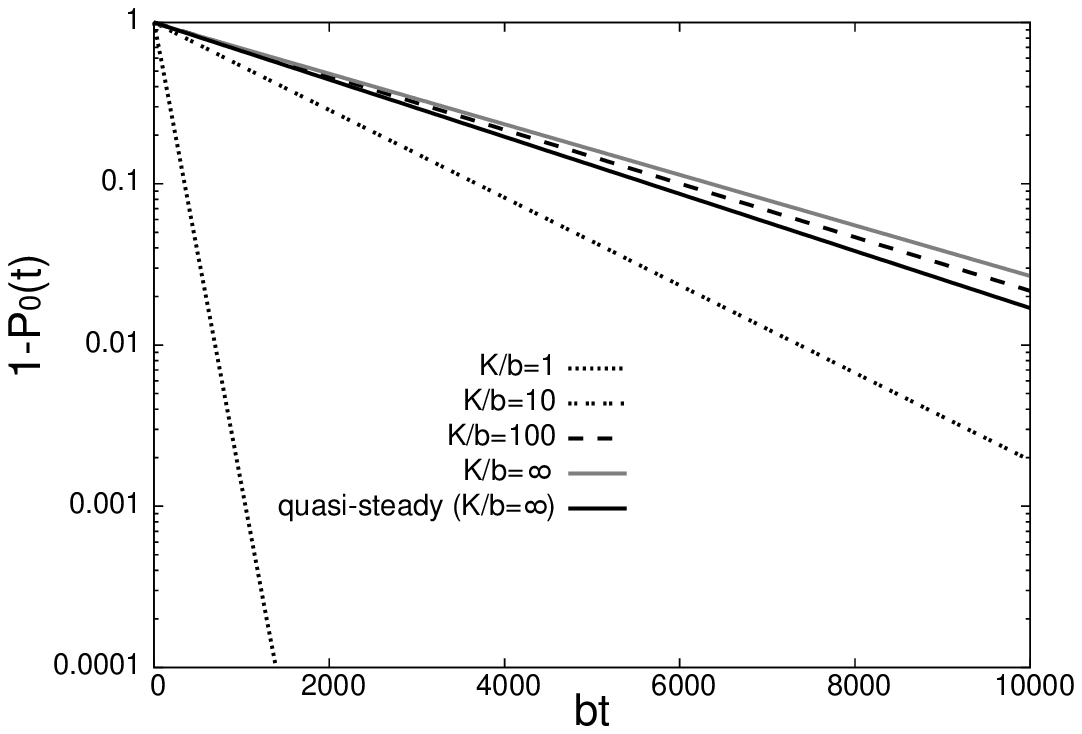}
\caption{The graphs of $1-p_0(t)$ for $a/b=10$, $r=1$, $\rho=0$, and $\epsilon=0$, for several values of $K/b$. The vertical axis is in log scale. The dashed lines are the results from the numerical integration of Eq.(\ref{master1}) with finite values of $K/b$. The gray solid line is the result for $K/b=\infty$, obtained from the numerical integration of Eq.(\ref{master4}). The black solid line is the result from the analytic expressions in Eqs.(\ref{expsl}) and (\ref{leakrate}).
}
\label{expf} % caption for the whole figure
\end{figure}

We have already argued in the previous section that the time-scales are more separated for larger values of $a/b-r$, where the short time-scale is the equilibration time of $m>0$ states, and the long time-scale is the one for the leakage to the $m=0$ state. To confirm this and to examine various properties of the leakage time, we performed numerical computation, using the method explained previously.  
The graphs of $p^{\rm st}_0 - p_0(t)=1-p_0(t)$ for $a/b=5$, $r=1$, $K/b=\infty$, $\rho=0$, and $\epsilon=0$, are shown in Figure \ref{expf5} with dashed lines for several initial conditions, where the vertical axis is in log scale. We indeed see that they form parallel straight lines for $bt \gtrsim 10$ , where $1-p_0(t) \lesssim 0.8$, confirming the single exponential form in Eq.(\ref{sexp}). The graphs of $1-p_0(t)$ for $a/b=10$, $r=1$, $\rho=0$, and $\epsilon=0$, are also plotted in Figure \ref{expf} for several values of $K/b$, where the single-exponential form is found for $1-p_0(t) \lesssim 0.9$ when $K/b \ge 1$, again indicating the time-scale separation. These results do not depend on initial probability distribution unless it is concentrated near $m=0$, which is again due to the time-scale separation~(Appendix \ref{exp}).

\begin{figure}
\includegraphics[width=\textwidth]{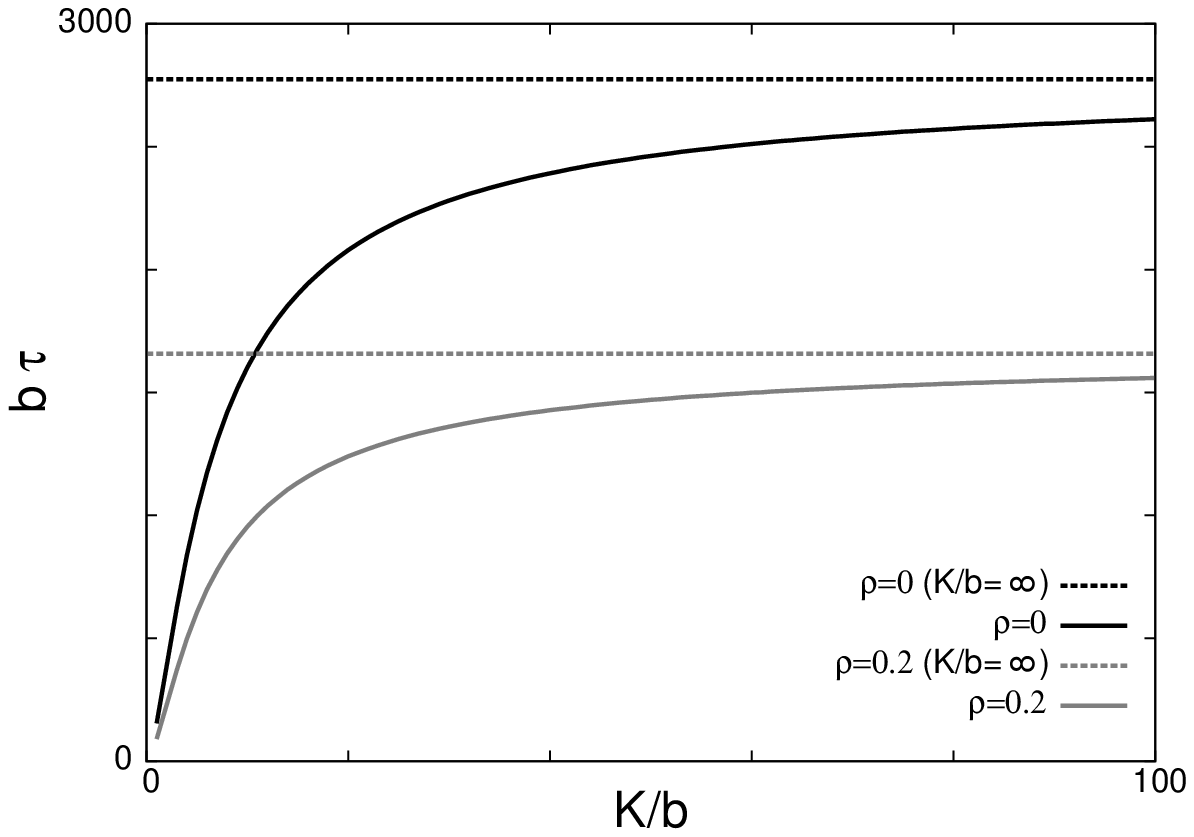}
\caption{The dimensionless mean leakage time $b \tau$ as the function of $K/b$, for $a/b=10$, $r=1$, and $\epsilon=0$. The dashed line shows the value at $K/b=\infty$. The black and the gray lines are for $\rho=0$ and $\rho=0.2$, respectively.
}
\label{tau} % caption for the whole figure
\end{figure}

\begin{figure}
\includegraphics[height=0.8\textheight]{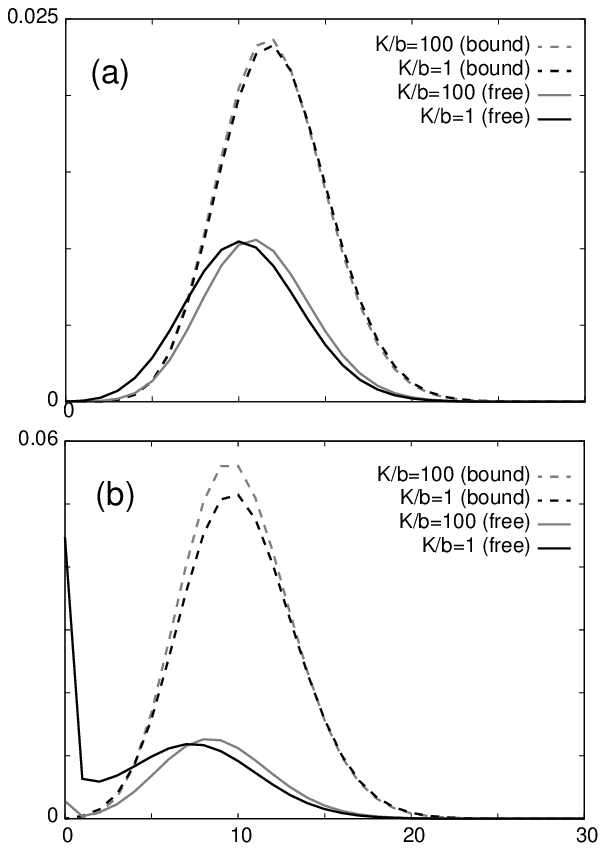}
\caption{The probability distributions $P(m,n,t)$ of bound ($n=1$, dashed lines) and free ($n=0$,  solid lines) modes (a) at $bt=1$ (b) and $bt=10$.   The distributions for $K/b=100$ and $K/b=1$ are compared, shown in gray and black lines, respectively. 	The other parameters are $b/a=10$, $r=1$, $\rho=0$, and $\epsilon=0$. For better visibility of the distributions of the free mode, those of the bound modes are scaled by 0.2 and 0.5 in figures (a) and (b), respectively. 
}
\label{fb} % caption for the whole figure
\end{figure}

\begin{figure}
\includegraphics[width=0.8\textwidth]{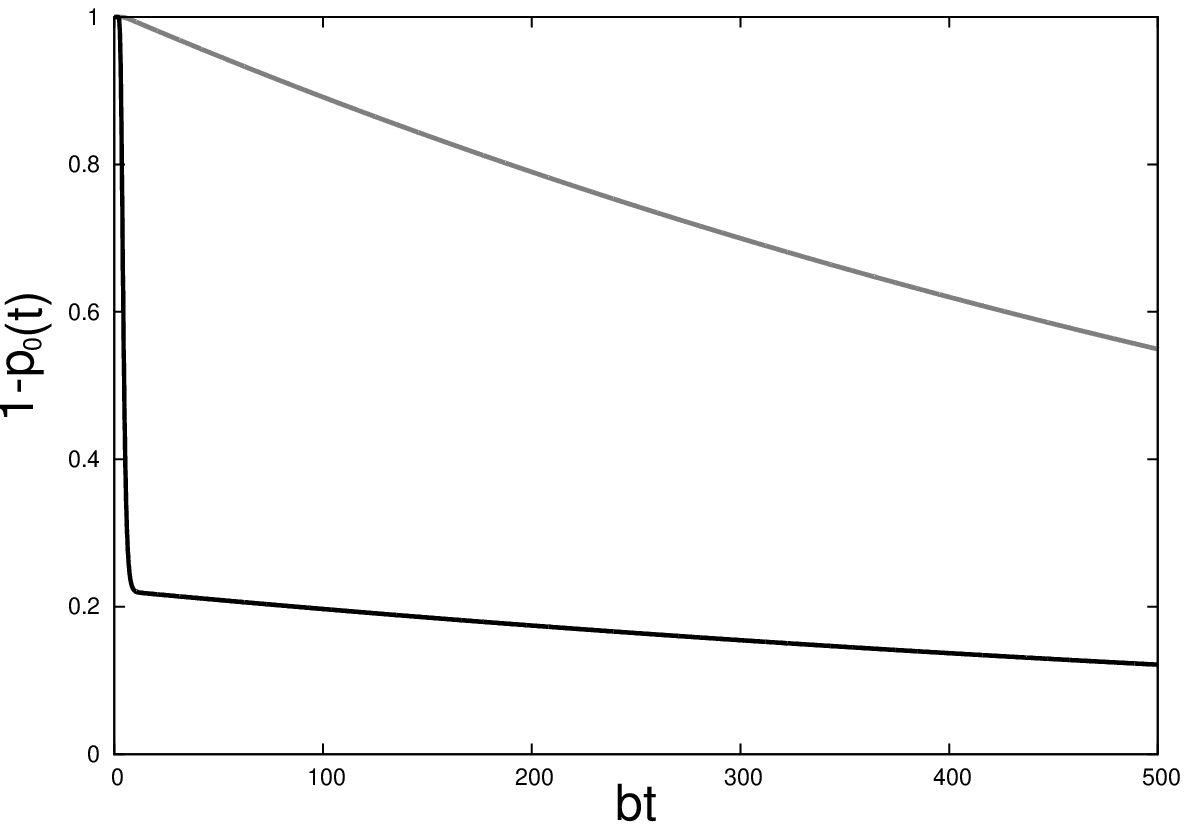}
\caption{The graphs of $1-p_0(t)$ for $\rho=0$, $\epsilon=0$, $a/b=100$, $r=0.4$, and $K=0.005$. The gray and the black lines are for the initial conditions $P(m,n,0)=\delta_{m,50} \delta_{n,1}$, and $P(m,n,0)=\delta_{m,50} \delta_{n,0}$, respectively.
}
\label{leak} % caption for the whole figure
\end{figure}

We also see that the increase of $K/b$ slows down the leakage to $m = 0$, which is also confirmed in the graph of the dimensionless mean leakage time $b\tau$ as a function of $K/b$ in Figure \ref{tau}, shown for both $\rho=0$ and $\rho=0.2$, with other parameters being the same as those in Figure \ref{expf}.  The faster leakage for a smaller value of $K/b$ is due to the free mode that flows straight down to $m = 0$ state without wasting time by making frequent transitions to the bound mode where the mean direction of flow is in the positive $m$ direction~(Fig.\ref{rate2}). This is even more evident from the separate snapshots of the time-dependent probability distributions for the bound and free modes in Figure \ref{fb}, where the behaviors for $K/b=1$ and $K/b=100$ are compared. We note in Figure \ref{tau} that increase of $\rho$ leads to the decrease of $\tau$ as is should, but the fact that it is an increasing function of $K/b$ remains unchanged. This feature was observed up to $\rho=1$ (Data not shown). Note that the fluctuation between the bound and the unbound mode is also a stochastic noise. From the results above, we see that the effect of this fluctuation is  qualitatively similar to that of the free protein number fluctuation, in that it enhances the leakage to the absorbing state. The behavior of $p_0(t)$ becomes highly dependent on initial conditions for $K/b \ll 1$, as in the case of a small value of $a/b-r$, due to the decoupling of the free and the bound mode: If the initial distribution is concentrated at the free mode, it is most probable that the proteins quickly get extinct before there is a chance for a protein to bind to the DNA, whereas if the initial distribution is concentrated on the bound state, it is most probable that protein number stays non-zero for some time before becoming extinct much later. The graphs of $1-p_0(t)$ are shown in Figure~\ref{leak} for $\rho=0$, $\epsilon=0$, $a/b=100$, $r=0.4$, and $K/b=0.005$, for the initial conditions $P(m,n,0)=\delta_{m,50} \delta_{n,1}$ (gray line), and $P(m,n,0)=\delta_{m,50} \delta_{n,0}$ (black line), respectively~\footnote{The states were truncated at $m=1000$. The probability at $m=1000$ remained below $2 \times 10^{-64}$ at all times.}.  We see that not only $p_0(t)$ depends on the initial condition, but also $1-p_0(t)$ is not even a single exponential for the second initial condition, indicating that the time-scale separation does not hold any more. These features can be most easily  understood by considering the extreme case of $K/b=0$ where analytic solution is available~(Appendix ~\ref{free}). It is obvious that a nonzero value of $K/b$ acts only as a perturbation if it is sufficiently small, and therefore the qualitative features of $K/b=0$ case are maintained. 

Note that the form of $p_0(t)$ obtained under the quasi-steady approximation, Eq.(\ref{expsl}), already has a single exponential form. This is because it is the solution of Eq.(\ref{appeq}) that is an approximation obtained under the assumption that the $m>0$ modes are equilibrated instantly. The assumption of instantaneous equilibration underestimates the leakage time, because actually it takes a finite time for a state with $m > 1$ to reach $m = 1$. However, it can be shown that $|\lambda_1| \to \tau_q^{-1}$ in the limit where $k_{1 \to 0}/k_{(m>0) \to (n>0)} \to 0$~(Appendix \ref{exp}). Therefore, the approximation is better for larger value of $a/b-r$. The graph of $1-p_0(t)$ obtained from the quasi-steady approximation, Eq.(\ref{expsl}), is also shown in Figures \ref{expf5} and \ref{expf}, where we indeed see that the leakage is faster than that of the exact solution at $K/b=\infty$, but  captures the exact behavior much better at $a/b-r=9$ where the time-scales are more separated, compared to $a/b-r=4$. 

In summary, for $a/b-r=4$ and $K/b=\infty$, the time-scales are sufficiently well separated in order for $p_0(t)$ to follow a single exponential form for $p_0(t) \gtrsim 0.2$, but not separated enough for $|\lambda_1|$ to be approximated by $\tau_q^{-1}$. For $a/b-r=9$ and $K/b=\infty$,  the time-scales are much better separated so that not only $p_0(t)$ follows a single exponential form for $p_0(t) \gtrsim 0.1$, but also $|\lambda_1| \simeq \tau_q^{-1}$. For $a/b-r=100$ and $K/b=0.005$, $p_0(t)$ does not follow single exponential form  in general, because the time-scales are not well separated.

\begin{figure}
\includegraphics[width=0.8\textwidth]{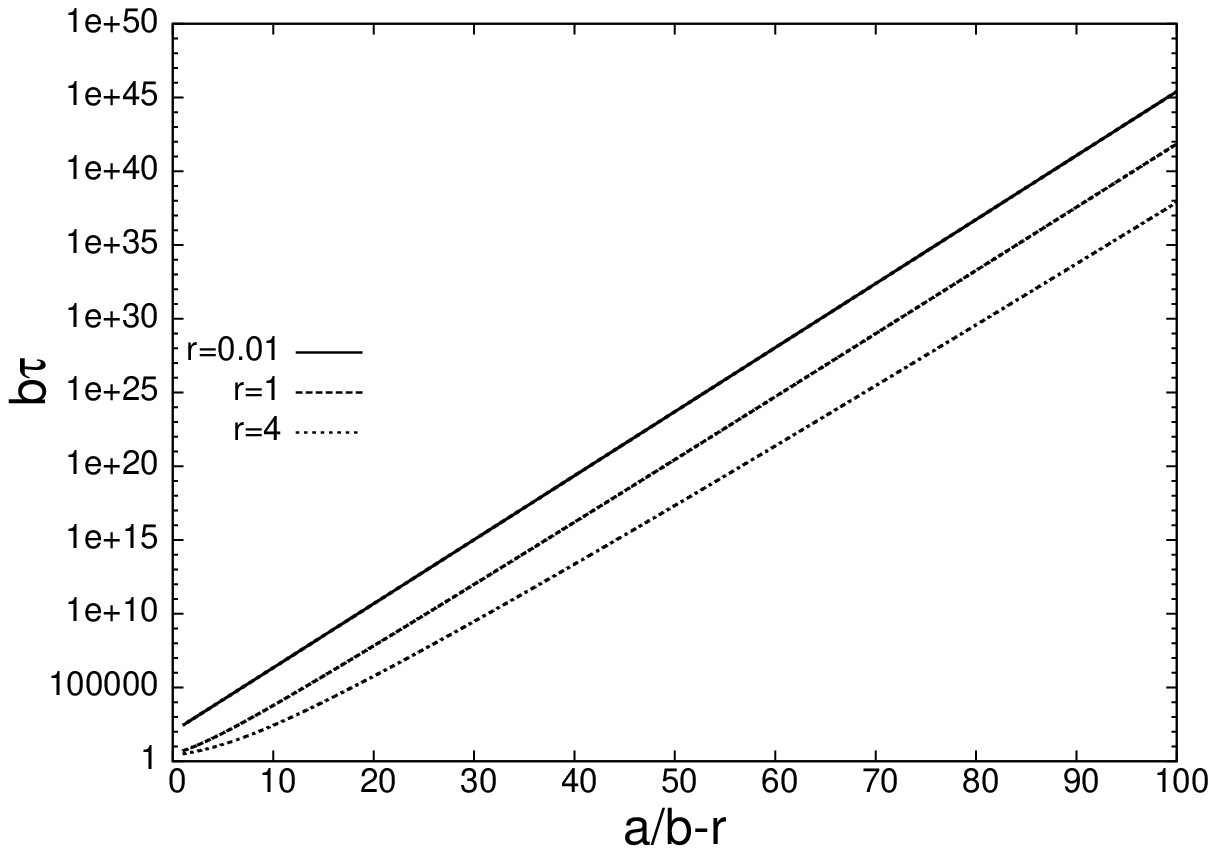}
\caption{The dimensionless mean leakage time $b \tau$ as the function $a/b-r$, for various values of $r$, for $K/b=\infty$, $\rho=0$, and $\epsilon=0$.}
\label{tauns} % caption for the whole figure
\end{figure}

Note that even for sufficiently large $K/b$, the probability distribution is dominated by the stable fixed point at $x=x_+$ only for $t \ll \tau \equiv |\lambda_1|^{-1}$. That is, the average behavior of the system follows the deterministic rate equation only at {\it early times}. For $t \gg \tau$, the probability distribution approaches the stationary one, and we have $p_m(t) \simeq \delta_{m,0}$. However, also note that larger the value of $m^*=a/b-r$, the smaller the stochastic noise, and hence better the deterministic approximation. In fact, the analytic form of $\tau_q$ for $K/b=\infty$ in Eq.(\ref{leakrate}) shows that it is an exponentially increasing function of $a/b-r$, as shown in Figure \ref{tauns} for several values of $r$,   and $\tau \to \infty$ in the limit of $a/b-r \to \infty$. That is, if the stochastic noise is very small, it takes a very long time for the probability distribution to make transition from the transient quasi-steady state to the true stationary one, and indeed the dynamics is well described by the deterministic rate equation for a long time duration. Because a large value of $r$ leads to the dominance of the free mode, it is obvious that the increase of $r$ speeds up the leakage, as shown in figure. 

\section{Effect of the baseline production}
When $\epsilon >0$, the numerator in Eq.(\ref{statsol}) does not diverge for $m=0$, and this expression is well defined for $m \ge 0$ with a nonzero value of $C$. Therefore,  $p_m(t)$ for $m>0$ does not vanish in the limit of $t \to \infty$, in contrast to the case of $\epsilon=0$. This is because $m=0$ is not an absorbing state any more. Similar situation is encountered in population dynamics, where influx of immigration plays the role of baseline production in the current model~\cite{meerson13}. However, the discussion for $\epsilon=0$ is still relevant when $\epsilon$ is sufficiently small, because the initial behavior of the time-dependent probability distribution is similar to that for $\epsilon=0$: The stable fixed point of the deterministic equation is the dominant state only at early times, and the occupation probability of the stable fixed point of the deterministic rate equation will be much smaller than $p_0(t)$ in the limit of $t \to \infty$, most of it concentrated at $m=0$. 

\begin{figure}
\includegraphics[width=\textwidth]{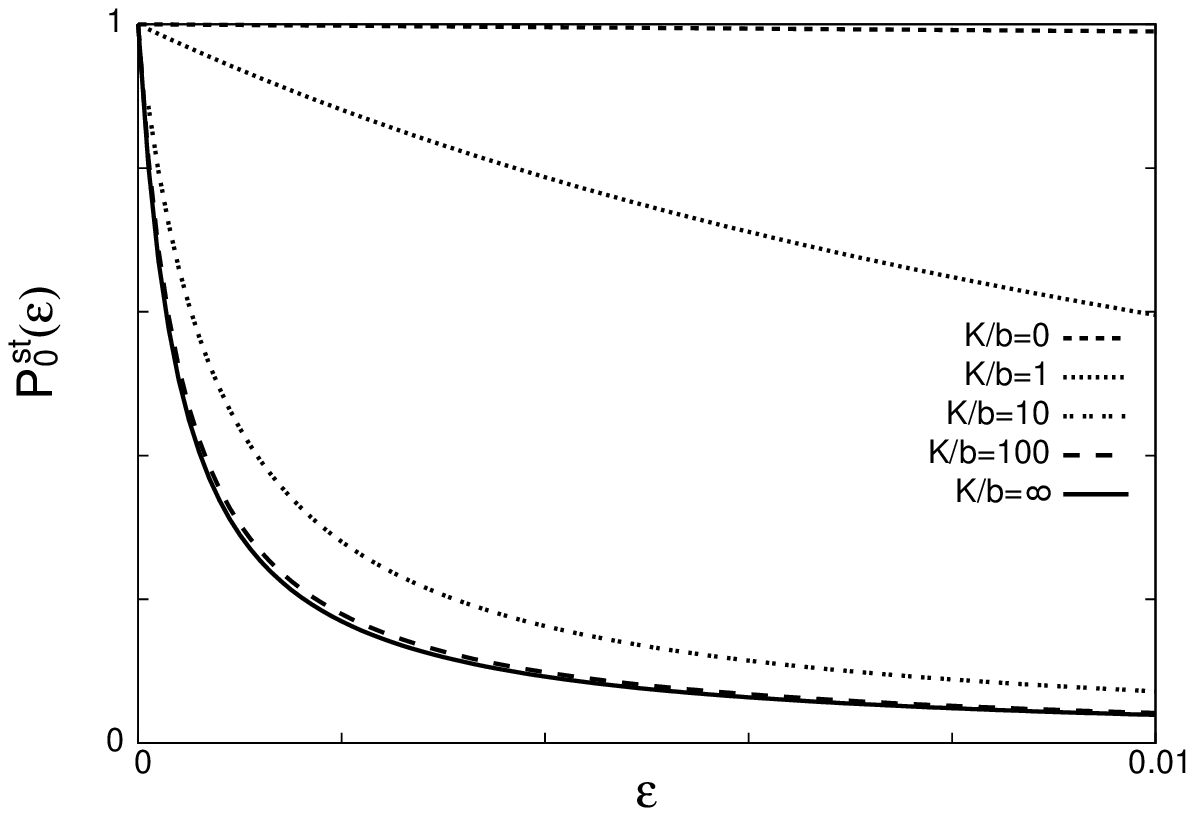}
\caption{The graphs of $p^{\rm st}_0(\epsilon)$ for $a/b=10$, $r=1$, and $\rho=0$, for various values of $K/b$.
}
\label{base} % caption for the whole figure
\end{figure}

By comparing Eqs.(\ref{statsol}) and (\ref{qsol}), we find that the functional form of $p^{\rm qs}_m (t)$ for $m \ge 1$ is approximately equal to that of $p^{\rm st}_m(\epsilon)$, up to the overall normalization constant, as long as $\epsilon$ is sufficiently small. Therefore, the stationary distribution for $\epsilon >0$ is approximately the same as the analytic form of the time-dependent distribution in Eq.(\ref{qsol2}) for $\epsilon=0$ at some time-point $t$. That is, we can find a pair of $t$ and $\epsilon$ satisfying $p^{\rm qs}_m (t) \simeq p^{\rm st}_m (\epsilon)$. For example, for $a/b=10$ and $r=1$, we find that $p(t=2000 b^{-1})$ for $K/b=100$ and $\epsilon=0$ shows a reasonably good agreement with $p^{\rm st} (\epsilon = 0.000035)$ for both $K/b=100$ and $K/b=\infty$~(Figure \ref{sim} (c)). The graph of $p^{\rm st}_0(\epsilon)$ is shown in Figure \ref{base} for several values of $K$, where we see that it is a monotonically decreasing function of $\epsilon$ as to be expected. The effect of $K/b$ on $p^{\rm st}_0(\epsilon)$ is similar to its effect on $p_0(t)$: a large  value  of $K/b$ hinders the flow of the probability to $m=0$~\footnote{The stationary distributions were obtained by using successive over-relaxation(SOR) algorithm~\cite{sor} that ensures fast convergence.}.

In summary, the baseline production ameliorates the effect of the stochastic noise in that $p^{\rm st}_m > 0$ for $m>0$, but for sufficiently small $\epsilon$, the qualitative behavior of the probability distribution is similar to that for $\epsilon=0$: at early times, the probability distribution converges to a quasi-steady distribution dominated by the stable fixed point, but the stable fixed point is almost erased in the limit of $t \to \infty$, although not completely destroyed. When $\epsilon$ is large enough so that the peak of $p^{\rm st}_m(\epsilon)$ around the stable fixed point is comparable to $p^{\rm st}_0(\epsilon)$ , we have a bistability driven by the stochastic noise in the limit of $t \to \infty$~\cite{fried}. For both of these cases, the deterministic rate equation describes the average behavior of the system only at early times. When $\epsilon$ is too large, then the effect of the baseline production dominates that of the stochastic noise in that $p_0^{\rm st}(\epsilon)$ is now smaller than the peak of $p^{\rm st}_m(\epsilon)$ at the stable fixed point. Then the deterministic rate equation description is valid throughout all the time scales, as long as the average behavior is concerned. Therefore, the effect of the baseline production is to oppose that of the stochastic noise. The stochastic noise and the baseline production have also been shown to exhibit opposite effects on the response to the the change of the production and/or the decay rates , the former and the latter favoring the binary and the graded responses, respectively~\cite{taba}. 

\begin{figure}
\includegraphics[width=0.8\textwidth]{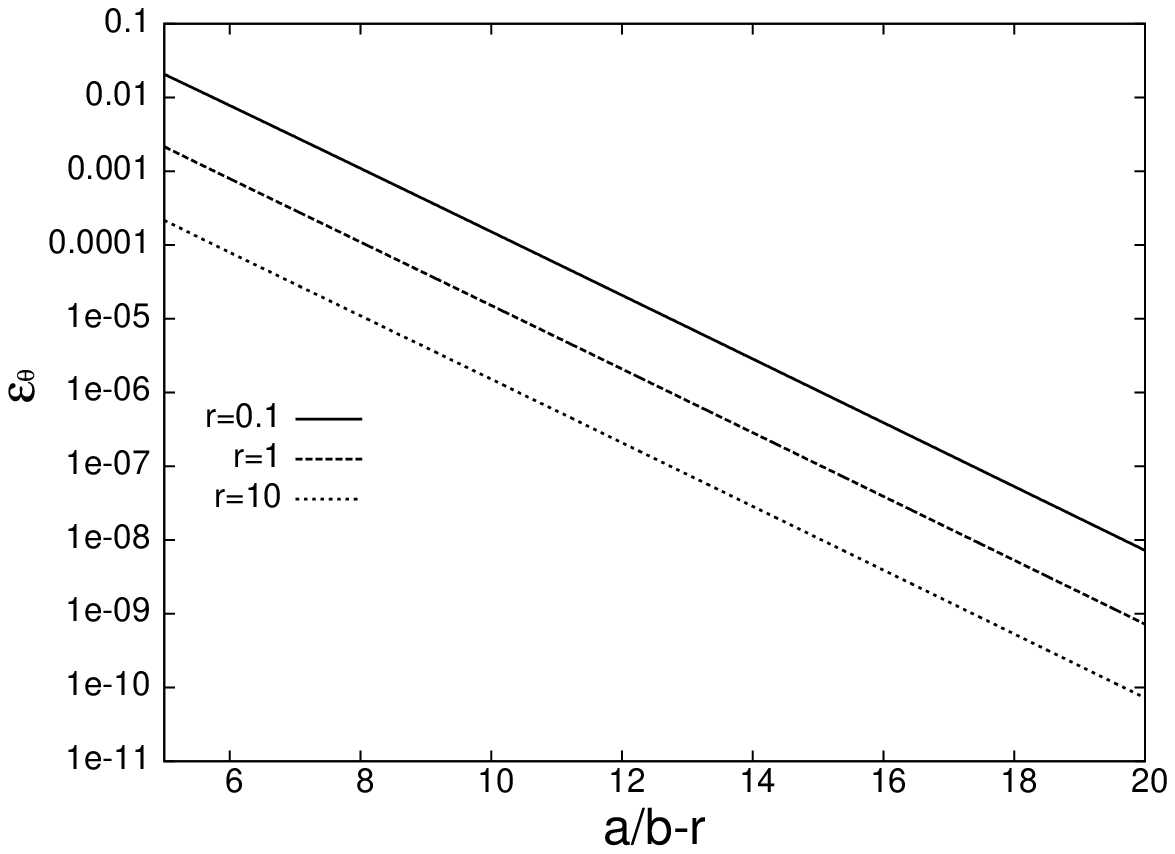}
\caption{The baseline production threshold $\epsilon_\theta$ as the function of $a/b-r$, for various values of $r$, for $K/b=\infty$ and $\rho=0$.}
\label{epsns} % caption for the whole figure
\end{figure}

The threshold value $\epsilon_\theta$, defined as the value of $\epsilon$ where the area under the two peaks are equal, is plotted in Figure~\ref{epsns} as the function of $a/b-r$,  for several values of $r$, for $\rho=0$, and $K/b = \infty$. 
We see that $\epsilon_\theta$ is an exponentially decreasing function of $a/b-r$. The fact that $\epsilon_\theta$ is a decreasing function of $a/b-r$ is to be expected: Because the effect of the stochastic noise and the baseline production oppose each other, larger (smaller) amount of baseline production is required to overcome the effect of the stochastic noise for larger (smaller) stochastic noise, corresponding to a smaller (larger) value of $a/b-r$. Also, for a larger value of $r$, the leakage effect is enhanced, and therefore more baseline production is required to resist such a leakage. As we will discuss in the next section, we can interpret the opposite effects of the stochastic noise and the baseline production in terms of the shift of the position of the unstable fixed point.

\section{Shift of the fixed points by stochastic noise}\label{pos}
We have seen in the context of the deterministic setting that the position of a fixed point gets shifted by the baseline production, and such a shift can remove the fixed point from the physical region. In the stochastic formalism, a fixed point turns into an extremum of the stationary distribution, and its position gets  shifted not only by the baseline production, but also by the stochastic noise~\cite{falk1,falk2}. To study this effect, and to see when the picture offered by the fixed points breaks down, we first go to the continuum limit where the chemical master equation for the stationary distribution turns into the Fokker-Planck equation of the form~\cite{kepler}~(Appendix \ref{FP})
\begin{equation}
-\partial_x \left( A(x) \pi^{\rm st}(x) \right) + \frac{1}{2} \partial_x^2  \left( B(x) \pi^{\rm st}(x) \right)=0. \label{sfp}
\end{equation}
where
\begin{eqnarray}
A(x) &=&  \frac{\tilde a (x + \tilde r \epsilon)}{x+\tilde r} -b x  +\frac{b x (1-\rho)}{\bar m (x+\tilde r)}\nonumber\\  
B(x) &=& \frac{1}{\bar m}  \left( \frac{\tilde a (x + \tilde r \epsilon)}{x+ \tilde r}+  b x \right). \label{FPn}
\end{eqnarray}
with $\tilde r = r /\bar m$ and $\tilde a = a /\bar m$. We note that a fixed point $x^*$ of the deterministic rate equation satisfies the equation $A(x^*)=0$ with $\bar m^{-1}=0$. The general solution of Eq.(\ref{sfp}) is~\footnote{There is another independent solution of the form $C B(x)^{-1} \exp\left(\int^x dz \frac{2A(z)}{B(z)}\right) \int^x dy \exp\left(-\int^y du \frac{2A(u)}{B(u)}\right)$, which is discarded by requiring that $\pi^{\rm st}(x)$  is normalizable. }
\begin{equation}
\pi^{\rm st}(x) = \frac{C}{B(x)} \exp\left(\int^x dz \frac{2A(z)}{B(z)}\right). \label{fpst}
\end{equation}
 Now let us consider the extremum of $\pi^{\rm st}(x)$ when $B(x)$ is very small. Taking derivative of $\pi^{\rm st}(x)$ with respect to $x$ and setting it to zero, we get
\begin{equation}
\frac{d \pi^{\rm st}(x)}{d x} = \frac{C}{B(x)} \exp\left(\int^x dz \frac{2A(z)}{B(z)}\right) \left[\frac{2 A(x)}{B(x)} - \frac{B'(x)}{B(x)}  \right] = 0,
\end{equation}
from which we get the equation for the extremum $x_m$:
\begin{equation}
A(x_m) - B'(x_m)/2 = 0.\label{fd}
\end{equation}
That is, we see that to the zeroth order of $\bar m^{-1}$, $x_m$ coincides with a fixed point $x^*$ of the deterministic rate equation, and the small stochastic noise acts as a perturbation that shifts the position of the $x_m$ with respect to $x^*$. To see whether $x_m$ is a local maximum or minimum of $\pi^{\rm st}(x)$, we compute the second derivative of $\pi^{\rm st}(x)$ at $x_m$:
\begin{eqnarray}
\frac{d^2 \pi^{\rm st}(x)}{d x^2}_{x=x_m} &=& \frac{C}{B(x)} \exp\left(\int^x dz \frac{2A(z)}{B(z)}\right) \left[\frac{2 A(x)}{B(x)} - \frac{B'(x)}{B(x)}  \right]^2_{x=x_m}\nonumber\\
&+& \frac{C}{B(x)} \exp\left(\int^x dz \frac{2A(z)}{B(z)}\right) \left[\frac{2 A'(x)}{B(x)} - \frac{2 A(x) B'(x)}{B(x)^2} - \frac{B''(x)}{B(x)}+ \frac{B'(x)^2}{B(x)^2} \right]_{x=x_m}\nonumber\\
&=& \frac{C}{B(x)} \exp\left(\int^x dz \frac{2A(z)}{B(z)}\right) \left[\frac{2 A'(x)}{B(x)}  - \frac{B''(x)}{B(x)} \right]_{x=x_m} \label{sd}
\end{eqnarray}
where Eq.(\ref{fd}) was used to derive the last line. From Eq.(\ref{sd}), we see that the sign of ${\pi^{\rm st}}''(x_m)$ is determined by $A'(x_m)-B''(x_m)/2$, which is $A'(x_*)|_{{\bar m}^{-1}=0}$ to the zeroth order in $\bar m^{-1}$. The result tells us that a stable (unstable) fixed point of the deterministic rate equation becomes a local maximum (minimum) of the steady-state probability distribution. There are two factors that modify the picture offered by the fixed point analysis. First, as shown earlier, the position shift of a fixed point may remove it from the physical region. Second, the probability distribution may possess an additional local maximum at the boundary of the physical region, $x=0$. This kind of local maximum is not related to a fixed point of the rate equation, because it is not obtained by taking the derivative to zero, and this new maximum may even dominate the behavior of the system. For the extreme case of $\epsilon=0$, we find that $B(x)^{-1} \sim 1/x \times {\rm constant}$ as $x \to 0$, and therefore the overall multiplicative constant in Eq.(\ref{fpst}) should vanish, leading to $\pi^{\rm st}(x) = \delta(x)$ as expected.

Now we analyze how perturbations of order ${\bar m}^{-1}$ and $\epsilon$ shift the positions of the extrema when $A(x)$ and $B(x)$ are given as in Eq.(\ref{FPn}). We have
\begin{eqnarray}
&&A(x_m)-\frac{B'(x_m)}{2}\nonumber\\
&=& \frac{\tilde a (x_m + \tilde r \epsilon)}{x_m +\tilde r} -b x_m  +\frac{b x_m (1-\rho)}{\bar m (x_m +\tilde r)} -\frac{1}{2 \bar m} \left[ \frac{\tilde a (x + \tilde r \epsilon)}{x+ \tilde r}+  b x \right]'_{x_m} \nonumber\\
&=& \frac{\tilde a (x_m + \tilde r \epsilon)}{x_m +\tilde r} -b x_m  +\frac{b x_m (1-\rho)}{\bar m (x_m +\tilde r)} - \frac{1}{2 \bar m } \left( \frac{\tilde a}{x_m+\tilde r} - \frac{\tilde a(x_m + \tilde r \epsilon)}{(x_m + \tilde r )^2}+  b  \right) = 0.  \label{fd1}
\end{eqnarray}
By multiplying Eq.(\ref{fd1}) by $x_m+\tilde r$, we obtain
\begin{equation}
-b x_m (x_m + \tilde r ) + \tilde a (x_m + \tilde r \epsilon)  - \frac{1}{2 \bar m } \left( \frac{\tilde a \tilde r (1 - \epsilon)}{x_m + \tilde r }+  b (\tilde r + (2 \rho -1) x_m)  \right) = 0 \label{fd2}
\end{equation}

Now, to find out the shift of $x_m$ to the leading order in ${\bar m}^{-1}$ and $\epsilon$, we make expansion $x_m = x^* + \delta x$, where $x^*$ is a fixed-point of the deterministic rate equation with $\epsilon=0$:
\begin{equation}
A(x^*)|_{\bar m^{-1} =\epsilon=0} = -b x^* + \frac{\tilde a x^*}{x^*+\tilde r} = 0 \label{fixed2}
\end{equation}
Eq.(\ref{fd2}) is now  expanded to the first order in $\bar m^{-1}$ and $\epsilon$, to obtain
\begin{equation}
(-2 b x^* -b \tilde r  + \tilde a) \delta x = -a \tilde r \epsilon  + \frac{1}{2 \bar m } \left( \frac{ \tilde a \tilde r }{x^* + \tilde r }+  b  (\tilde r + (2 \rho -1 ) x^*) \right) +O({\bar m}^{-2},\epsilon^2,{\bar m}^{-1} \epsilon), \label{fd3}
\end{equation}
from which we obtain that
\begin{equation}
 \delta x = (-2 b x^* -b \tilde r  + \tilde a)^{-1}\left[-\tilde a \tilde r \epsilon  + \frac{1}{2 \bar m } \left(  \frac{\tilde a \tilde r }{x^* + \tilde r }+  b  (\tilde r + (2 \rho -1 ) x^*) \right) \right]. \label{exfd}
\end{equation}
When $x^*=0$, we get
\begin{equation}
\delta x = (\tilde a -b \tilde r  )^{-1}\left[-\tilde a \tilde r \epsilon  + \frac{1}{2 \bar m } \left( \tilde a +  b  \tilde r \right) \right].
\end{equation}
If $\tilde a > b \tilde r$ so that $x^*=0$ is a local minimum (unstable fixed point), then $\delta x > 0$ if $\bar m^{-1} ( \tilde a +  b  \tilde r )  > 2 \tilde a \tilde r \epsilon$ and $\delta x < 0$ otherwise.  The stable fixed point also tends to get shifted in the opposite directions by the stochastic noise and the baseline production(~Appendix \ref{shiftx}). 

\begin{figure}
\includegraphics[width=0.8\textwidth]{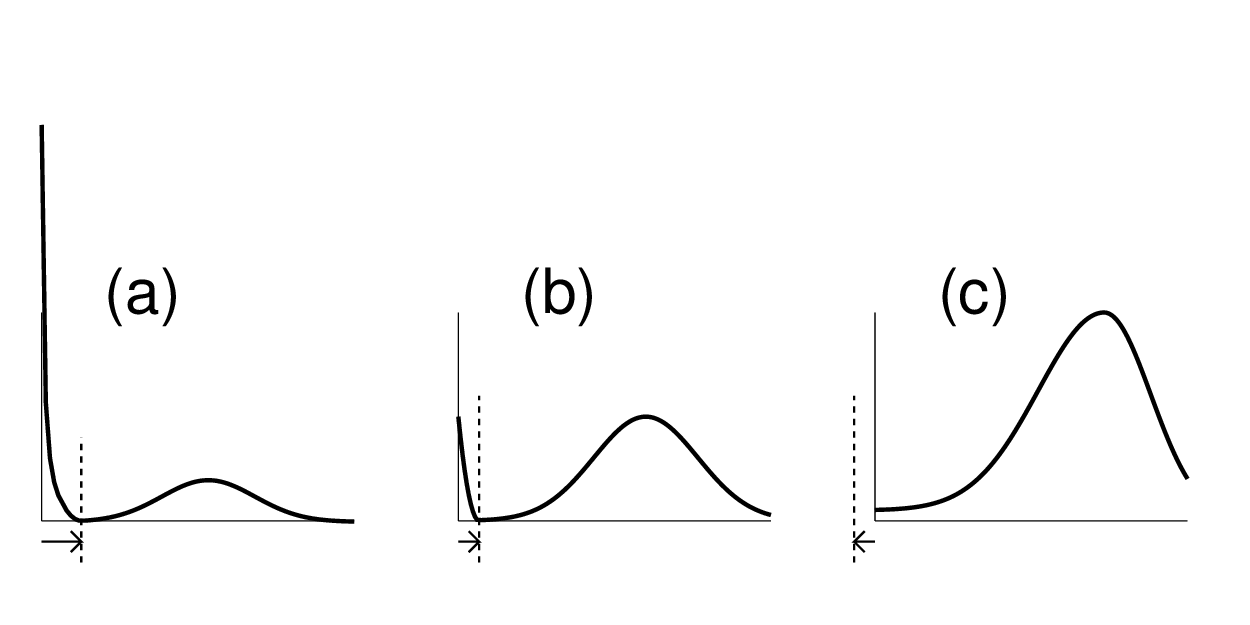}
\caption{The shift of the minimum of the stationary distribution. In each figure, the tail of the arrow is located at the unstable fixed point $x_-=0$ of the deterministic rate equation with no baseline production. (a,b) The local minimum is shifted rightward, when the effect of the stochastic noise dominates over that of the baseline production. (a) For large noise and the corresponding large shift of the unstable fixed point, a divergence occurs at $x=0$, leading to the effective erasure of the non-zero stable fixed point for $t \to \infty$. (b) For small magnitude of noise, probability at $x=0$ becomes comparable to the peak near the non-zero stable fixed point, leading to the noise-induced bistability for $t \to \infty$. (c) When the effect of the baseline production dominates that of the stochastic noise, the local minimum is shifted leftward. The local minimum disappears into the unphysical region of $x<0$, and the probability distribution is dominated by the global maximum near the non-zero stable fixed point, even for $t \to \infty$. 
}
\label{shift} % caption for the whole figure
\end{figure}

In summary, the opposite effects of the stochastic noise and the baseline production on the stationary distribution can be  interpreted in terms of the position shift of the unstable fixed point: The former and the latter tends to shift the position of the unstable fixed point into the positive and the negative directions, respectively. The erasure of stable fixed point by stochastic noise, as well as the noise-driven bistability, can be understood in terms of the position shift of the unstable fixed point at $x=0$~(Figure \ref{shift}). When the effect of the stochastic noise is larger than that of the baseline production, the position of the unstable fixed point $x=0$ shifts to the region of positive $x$, and $x=0$ becomes the local maximum of the probability distribution without vanishing derivative, which may even dominate the behavior of the stationary state if the shift is sufficiently large.

\section{Numerical estimates using biological parameters}
\begin{figure}
\includegraphics[width=0.8\textwidth]{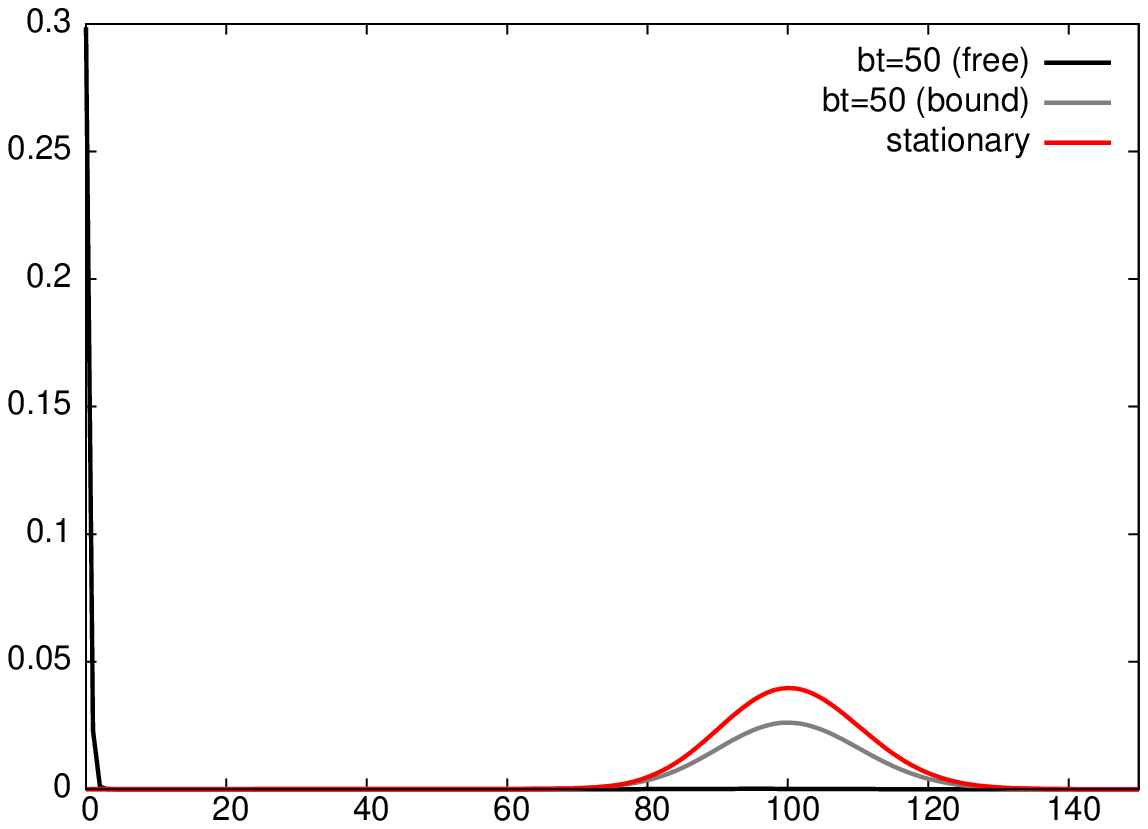}
\caption{The probability distribution at $bt=50$, for the initial condition $P(m,n,0)=\delta_{m,0} \delta_{n,0}$ and the parameters $\rho=0$, $a/b=100$, $r=0.4$, $K=0.3$, and $\epsilon=0.001$. The distributions for the free ($P(m,0,t)$)  and the bound mode ($P(m,1,t)$) are shown as black and the gray lines, respectively. The stationary distribution is also shown in red line. The free mode makes negligible contribution to the stationary distribution  and therefore $p^{\rm st}_m \simeq P^{\rm st} (m-1,1)$. Probability distribution at $bt=10^6$ for $\rho=0$, $a/b=100$, $r=0.4$, $K=0.3$, and $\epsilon=0$, is indistinguishable from $p^{\rm st}_m$, as long as the initial state is away from $m=0$.
}
\label{final} % caption for the whole figure
\end{figure}

Most of the parameters used below are for the  Lac system of the bacteria {\sl Escherichia Coli}. Although Lac system is not a positive feedback loop, these parameters are used just for the order of magnitude estimate\footnote{The authors thank an anonymous referee for providing the parameters.}. 
The maximal production rate of the {\sl E. Coli} is $a \sim 100\ {\rm min}^{-1}$ (BNID 100738)\footnote{The ID number of BioNumber Database~\cite{BN}.}, using the data for the protein LacY and LacZ~\cite{alon,santi}.  Even for most unstable protein in {\sl E. coli}, the degradation rate is $b \sim 1\ {\rm min}^{-1}$ ~\cite{mazu,dg1,dg2,dg3,dg4} (BNID 109921), giving the estimate of 
\begin{equation}
a/b \gtrsim 100. \label{one}
\end{equation}
To estimate $r$, we use  the values of binding constant $k_0 V = 0.0027 ({\rm s\  \rm nM})^{-1}\sim 0.003 ({\rm s\  \rm nM})^{-1}= 0.003 \ {\rm s}^{-1} \times (10^{-9}\ {\rm mol})^{-1} \times 10^{-3}\ {\rm  m}^3 $~\cite{bu1} and unbinding constant $k_1 = 0.0023   ({\rm s})^{-1} \sim 0.002   ({\rm s})^{-1}$~\cite{bu2} of {\sl E. coli} {\sl lac} repressor (LacL) ~\cite{snch} (BNID 106521), as well as the volume of  {\sl E. Coli}, $V \sim 1 \mu m^3 = 10^{-18}\  {\rm m}^3$ \cite{vol1,vol2} (BNID 114924, 114925), to get
\begin{eqnarray}
k_0/b &\sim& \frac{0.003  ({\rm s\  mol})^{-1}  \times 10^6 \ {\rm m}^3}{6 \times 10^{23}\  {{\rm molecules}\ {\rm mol}^{-1}}  \times 10^{-18} {\rm m}^3} \times 60 \  {\rm s} = 0.3 \ {\rm molecule}^{-1}, \nonumber\\ 
k_1/b &\sim&  0.002 \  {\rm s}^{-1} \times 60 \ {\rm s}  = 0.12,  \nonumber\\ 
r &=& k_1/k_0 \sim 0.4. \label{two}
\end{eqnarray}
For simplicity we assume $\rho=0$ throughout the estimates. The numerical computation for the parameters given above cannot be conducted long enough to compute $\tau$, due to the accumulation of numerical errors. However, the probability distribution remains peaked around $m\sim 100$ and shows no sign of leakage to $m=0$, up to $10^6\ {\rm min}$~(Fig.~\ref{final}, red line), regardless of the initial condition. This indicates that $\tau \gg 10^6\ $min. Considering the fact that the cell generation time of  {\sl E. Coli} is $T_{\rm cycle} \sim 100\ {\rm min}$~\cite{cyc} (BNID 105065), we see that
\begin{equation}
\tau \gg T_{\rm cyc}.
\end{equation}
These results suggest that  the leakage to zero-protein state will be unobservable in real biological system even if there is no baseline production.

In reality, $\epsilon>0$ except for artificially engineered systems~\cite{zero1,zero2,zero3}. 
For Lac promotor, we have $\epsilon=10^{-3}$~\cite{base} (BNID 102075). With other parameters given as above, this amount of baseline production is sufficient to dominate over the effect of the stochastic noise, as shown in the stationary distribution that is peaked around $m \sim 100$, with no trace of peak near $m=0$~(Fig.~\ref{final}).  The effect of the baseline production is more important than that of the large value of $\tau$, because it helps the system to start the positive feedback loop even if there is no protein in the beginning. Even if use the initial condition $P(m,n,0)=\delta_{m,0}\delta_{n,0}$, the amount of the baseline production above is sufficient to let the system quickly escape from the state $(m,n)=(0,0)$. Only 30\% of the probability remains at $m=0$ at $t = 50$\  min, as shown in Figure~\ref{final}. Therefore, from these results, we expect that for wild-type genetic regulatory networks, the baseline production will restore the non-zero stable fixed point as the dominant peak of the stationary distribution.

\section{Discussion}
It is a well-known fact that stochastic noise modifies the picture provided by the deterministic rate equation. A representative example is the conversion of the stable fixed point into a transient peak of the probability distribution and its complete removal from the stationary distribution. Although this phenomenon has been extensively studied in the context of population dynamics and epidemics, it has been seldom discussed for the models of gene-regulatory networks.     

In this work, we performed quantitative analysis of transient dynamics of the simplest auto regulatory genetic circuit with positive feedback, both numerically and analytically. We found that as long as the magnitude of the baseline production is sufficiently small compared to that of the stochastic noise, the unique stable fixed point turns into a dominant peak of the transient, quasi-steady distribution, instead of the true stationary state. In the extreme case of vanishing baseline production, the trace of the stable fixed point is completely erased from the stationary distribution due to an absorbing state. However, for very small stochastic noise, the probability distribution is dominated by the stable fixed point for a very long time duration. In fact, we find that the leakage time is an exponentially increasing function of the inverse square of the relative fluctuation, $a/b-r$~(Fig.\ref{tauns}). This clarifies the true meaning of the statement that the chemical master equation is well approximated by the deterministic rate equation when the stochastic noise is small. For a given time, the deterministic rate equation becomes a better approximation of the system as the noise is reduced. Also, the time duration for which deterministic description is valid becomes longer as the noise is reduced. The  contradiction between the stochastic and the deterministic equation appeared only because we took the $t \to \infty$ limit first. Considering that the biological processes occur within a finite time duration, the transient behavior may more be biologically relevant than the true stationary distribution of the chemical master equation. The importance of the transient behavior has also been emphasized for the stochastic decision process in $\lambda$-phage system~\cite{weitz}\footnote{The main difference from the current result is that for the $\lambda$-phage decision circuit, the transient and the stationary behaviors of the deterministic and stochastic equations coincide, whereas for the simple auto-regulatory model considered here, the stationary behavior of the deterministic equation corresponds to the transient one of the stochastic equation.}. 

In reality, there is always a small amount of baseline production from an inactive gene, except for artificially engineered systems~\cite{zero1,zero2,zero3}. Because the baseline production removes the absorbing state, its effect is opposite to that of the stochastic noise. The magnitude of the baseline production relative to that of the stochastic noise determines the relative dominance of the non-zero stable fixed point relative to the peak at the zero-protein state, in the stationary distribution. In fact, we find that the baseline production rate required for overcoming the stochastic effect is an exponentially decreasing function of the inverse square of the relative fluctuation, $a/b-r$~(Fig.\ref{epsns}). We also showed that the opposite effects of the stochastic noise and that of the baseline production can be interpreted in terms of the position shift of the unstable fixed point. 

The order of magnitude estimates using biological parameters suggest that for a real gene regulatory network, the stochastic noise is sufficiently small so that not only is the leakage time much larger than biologically relevant time-scales, but also the effect of the baseline production completely dominates over that of the stochastic noise. Therefore, the wild-type gene-regulatory networks seem to be protected from the catastrophic rare event of protein extinction, by both of these effects.

%%%%%%%%%%%%%%
\section{Acknowledgement} 
This work was supported by the National Research Foundation of Korea, funded by the Ministry of Education, Science, and Technology (NRF-2017R1D1A1B03031344). We thank Hyunggyu Park, Seung Ki Baek, Changbong Hyeon, Kingshuk Ghosh, and Nam Ki Lee for useful comments and discussions.

%This is a test~\cite{gil1}.
%\bibliographystyle{plain}
%\bibliographystyle{unsrt}
%\bibliographystyle{aipauth4-1}
%\bibliographystyle{jabbrv_siam}
%\begin{thebibliography}{}
\bibliography{stoch}
%\end{thebibliography}

\appendix
\section{Demonstration of vanishing stationary probability of non-zero protein numbers in other models of gene regulatory network}\label{other}
The delta function form of the stationary distribution, Eq.(\ref{deltaftn}), is due to the fact that the zero-protein state is an absorbing state of the system. Therefore, the stationary state is either Kronecker delta function or Dirac delta function concentrated at the zero-protein state, as long as there is no baseline production from the inactive gene, and the result does not depend on a  specific form of the noise.  We will consider several models in the literature below.

Let us consider the model  defined by the continuous master equation~\cite{fried,taba}
\begin{equation}
\frac{\partial p(x)}{\partial t}  = \frac{\partial}{\partial x} [\gamma_2 x p(x)] + \frac{k_1}{b} \int_0^x d x' \exp(-(x-x')/b)\left(\frac{1  + \epsilon c x'^H}{1  + c x'^H} \right)p(x') - p(x)
\end{equation} 
where $x$ is the concentration of the protein. Here, because the decay and the production terms are given by the first-order derivative and the integral terms, respectively, this model has less degradation noise and more production noise compared to the Fokker-Planck description, where both the decay and the production are described by the second-order derivatives. The stationary solution is given by~\footnote{The solutions given in ref.\cite{fried} and ref.\cite{taba} are slightly different, but we adopt the convention of the latter which is more natural. The difference disappears for $\epsilon \to 0$. }
\begin{equation}
p^{\rm st}(x) = A x^{\alpha-1} e^{-x/\beta} (1+c x^H) ^{\alpha (\epsilon-1)/H}
\end{equation}
The positive regulation corresponds to $H<0$, and in this case $1+c x^H  \to c x^H$ as $x \to 0$.  Therefore, $p(x) \to A x^{\alpha\epsilon - 1}$ as $x \to 0$, and $\int_0^\infty dx p(x)$ diverges if $\epsilon=0$ and $A>0$.  In fact, the normalization $\int dx  p(x)=1$ requires that $A \simeq \alpha \epsilon \to 0$ as $\epsilon \to 0$. Therefore, $p(x)=0$ for $x>0$, and consequently $p(x) = \delta _+(x)$ for $\epsilon=0$, where the distribution $\delta_+(x)$ is defined by the property that $\delta_+(x)=0$ for $x>0$ and $\int_0^{\infty} dx \delta_+(x) = 1$. 

Next we consider a discrete model where the proteins form dimer in the bulk and then bind to either DNA or RNA for positive regulation~\cite{aquino}.  For the transcriptional regulation with fast mRNA dynamics, we have the stationary solution of the form
\begin{equation}
p^{\rm st}_n = \frac{r p^{\rm st} _0}{n} \prod_{i=1}^{n-1} \left( r \frac{f(i)}{i} + \frac{\mu_p-1}{\mu_p}\right) \label{di1}
\end{equation}
for $n>0$,  where
\begin{equation}
f(n) \equiv \sum_{j \ge 0} \frac{1 + \rho k j}{1 + kj}   \frac{(\lambda n_2 (n))^j}{j!} \exp(-\lambda n_2 (n))
\end{equation}
with
\begin{equation}
n_2(n) = \frac{n}{2} + a^2 - a \sqrt{n+a^2}.
\end{equation}
The zero baseline production corresponds to the limit of $r \to 0$ and $\rho \to \infty $ with finite value of  $r \rho$ which is proportional to the transcription rate from the active DNA .  Because $r \frac{f(i)}{i}$ in the parenthesis of Eq.(\ref{di1}) remains finite in this limit, $p^{\rm st}_n$ for $n>0$ all vanish due to the extra $r$ in the front of the right-hand side of  Eq.(\ref{di1}), leading to $p^{\rm st}_n = \delta_{n,0}$ due to the normlization.

In the continuum limit, Eq.(\ref{di1}) is approximated as
\begin{equation}
p^{\rm st}(x) = A_c x^{-1} e^{-x/\tilde \mu_p} \exp( r \int_c^x du \tilde f(u)/u). \label{co1}
\end{equation}
where 
\begin{equation}
\tilde f(x) \equiv \sum_{j \ge 0} \frac{1 + \rho k j}{1 + kj}   \frac{(\lambda x_2 (x))^j}{j!} \exp(-\lambda x_2 (x))
\end{equation}
with 
\begin{equation}
x_2(x) = \frac{x}{2} + \lambda a^2 - \sqrt{\lambda} a \sqrt{x+\lambda a^2}
\end{equation}
Because the zero baseline production corresponds to $r \to 0$ with $r \rho$ finite, we have 
\begin{equation}
r f(x) =  \sum_{j \ge 0} \frac{ r \rho k j}{1 + kj}   \frac{(\lambda x_2 (x))^j}{j!} \exp(-\lambda x_2 (x)),
\end{equation}
and since $x_2(x) \to x/2$ as $x \to 0$, we have
$r f(x) \propto x$ for small $x$ in the case of zero baseline production, and consequently $r  \int_c^x du f(u)/u$ is non-zero and finite. Therefore, from Eq.(\ref{co1}) we see that $p^{\rm st}(x) \propto x^{-1}$ as $x \to 0$, and again we see that the integral of  $p^{\rm st}(x)$ diverges unless $A_c=0$. Therefore, we again see that $p(x) = \delta_+(x)$. When there is a non-zero baseline production,  $r f(x) \to r$ as $x \to 0$, and  $r  \int_c^x du f(u)/u  = r \ln (x/c)$. Therefore  $p^{\rm st}(x) \propto x^{r-1}$ as $x \to 0$ so the integral of $p^{\rm st}(x)$ remains finite.

The discrete and continuous solutions for the transcriptional regulation under the fast protein dynamics, as well those under translational regulations, have similar forms as the ones presented above, so they can be shown to reduce to Dirac delta and Kronecker delta functions respectively in the absence of the baseline production, following the same logic as above. 

\section{Derivation of the reduced master equation (\ref{master4}) in the limit of fast equilibration of DNA.}\label{mdv}
We redefine the parameters and the variables in Eq(\ref{master1}):
\begin{eqnarray}
K &\equiv& k_0, \quad r \equiv k_1/k_0 \nonumber\\
p_m &\equiv& P (m,0) + P(m-1,1) \nonumber\\
\xi_m &\equiv& \frac{m  P(m,0) - r P(m-1,1)}{m  + r},\label{redef}
\end{eqnarray}
where the time index is suppressed for notational simplicity. $P(m,n)$ are then expressed in terms of $p_m$ and $\xi_m$ as
\begin{eqnarray}
P(m,0) = \frac{r}{m+r} p_m + \xi_m \nonumber\\
P(m-1,1) = \frac{m}{m+r} p_m - \xi_m \label{redef2}
\end{eqnarray}
By substituting Eq.(\ref{redef2}) into Eq.(\ref{master1}), we obtain
\begin{eqnarray}
\dot p_m &=& \dot P(m,0) + \dot P(m-1,1) \nonumber\\
&=& \epsilon a P(m-1,0)  - \epsilon a P(m,0) + aP(m-2,1) -aP(m-1,1)   \nonumber\\
&& +  bP(m+1,0)(m+1) - bP(m,0) m +  bP(m,1)m  - bP(m-1,1)(m-1)  \nonumber\\&&+ \rho b P(m,1) - \rho b P(m-1,1),\nonumber\\
&=& a(\frac{m-1}{m+r-1} p_{m-1} - \xi_{m-1} - \frac{m}{m+r} p_{m} + \xi_{m}) + \epsilon a (\frac{r}{m+r-1} p_{m-1} + \xi_{m-1}   - \frac{r}{m+r} p_m - \xi_m ) \nonumber\\
&& +  b (m+1) (\frac{r}{m+r+1} p_{m+1} + \xi_{m+1})  - b m (\frac{r}{m+r} p_{m} + \xi_{m}) \nonumber\\
&&+  b (m + \rho) (\frac{m+1}{m+r+1} p_{m+1} - \xi_{m+1}) - b(m-1+ \rho) (\frac{m}{m+r} p_m - \xi_m)  \nonumber\\
&=&   \frac{(m-1 + \epsilon  r) a     }{m-1  + r} p_{m-1} -  \frac{ (m + \epsilon  r) a }{m  + r} p_{m} + \frac{b (m+1) (r + m + \rho)}{m+1  + r} p_{m+1} -  \frac{ m b  (r + m-1 + \rho)}{m  + r} p_{m} \nonumber\\
&&- a (1 - \epsilon) \xi_{m-1} + (a(1 - \epsilon)-b) \xi_{m} +   b \xi_{m+1}  \label{split1}
\end{eqnarray}
and
\begin{eqnarray}
\dot \xi_m &=& \frac{m  \dot P(m,0) - r \dot P(m-1,1)}{m  + r} \nonumber\\
&=& \frac{m}{m  + r} \Big(-k_0 P(m,0) m + k_1 P(m-1,1) + \epsilon a P(m-1,0)  - \epsilon a P(m,0) +b P(m+1,0)(m+1) \nonumber\\
&&- b  P(m,0) m + \rho b P(m,1) \Big) - \frac{r}{m  + r} \Big(k_0 P(m,0) m - k_1 P(m-1,1) +  a P(m-2,1)  \nonumber\\
&& - a P(m-1,1)+b P(m,1)m - b  P(m-1,1) (m-1+\rho) \Big)\nonumber\\
&=& (-k_0 m - \frac{\epsilon a m}{m+r} - \frac{b  m^2}{m+r} ) P(m,0)+ (k_1 + \frac{a r}{m+r} + \frac{b (m-1+\rho) r}{m+r})P(m-1,1)\nonumber\\
&&+ \frac{\epsilon a m}{m+r} P(m-1,0) +\frac{b m (m+1)}{m+r} P(m+1,0) - \frac{ a r }{m+r} P(m-2,1) + \frac{ b m (\rho  -r)}{m+r} P(m,1) \nonumber\\
&=& (-k_0 m - \frac{\epsilon a m}{m+r} - \frac{b  m^2}{m+r} ) (\frac{r}{m+r} p_m + \xi_m) + (k_1 + \frac{a r}{m+r} + \frac{b (m-1 + \rho) r}{m+r})( \frac{m}{m+r} p_m - \xi_m )\nonumber\\
&&+ \frac{\epsilon a m}{m+r}(\frac{r}{m+r-1} p_{m-1} + \xi_{m-1}) +\frac{b m (m+1)}{m+r}(\frac{r}{m+r+1} p_{m+1} + \xi_{m+1}) \nonumber\\
&&- \frac{ a r }{m+r} ( \frac{m-1}{m+r-1} p_{m-1} - \xi_{m-1}) + \frac{ b m (\rho - r )}{m+r}  (\frac{m+1}{m+r+1} p_{m+1} - \xi_{m+1})\nonumber\\
&=& -K (m  + r) \xi_m - \left (\frac{a (r+\epsilon m)} {m  + r} + \frac{b (m^2  + (m-1+\rho) r)}{m+r} \right) \xi_m + \left( \frac{a r+\epsilon a m} {m  + r} \right)\xi_{m-1} \nonumber\\
&& + \frac{b m(m  + r + 1-\rho)}{m  + r}  \xi_{m+1}   +  \frac{m (a r(1-\epsilon) +(\rho-1) b)}{(m  + r)^2} p_m   -  \frac{ a r(m-1-\epsilon m)}{(m  + r)(m  + r-1)} p_{m-1}  \nonumber\\
&&+ \frac{ b m \rho (m+1)}{(m  + r)(m  + r-1)} p_{m+1}.  \label{split2}
\end{eqnarray}
From  Eq.(\ref{split2}), we get
\begin{eqnarray}
\xi_m &=& - \frac{\dot \xi_m}{K (m  + r)}  -  \frac{a (r + \epsilon m) + b (m^2  + (m-1+\rho) r)} {K(m  + r)^2}\xi_{m} + \left( \frac{a r+\epsilon a m} {K(m  + r)^2} \right)\xi_{m-1} \nonumber\\
&& + \frac{b m(m  + r + 1-\rho)}{K(m  + r)^2}  \xi_{m+1}    +  \frac{m (a r(1-\epsilon) + (\rho - 1) b)}{K(m  + r)^3} p_m  -  \frac{ a r(m-1-\epsilon m)}{K (m  + r)^2(m  + r-1)} p_{m-1}\nonumber\\ 
&& + \frac{ b m \rho (m+1)}{K (m  + r)^2 (m  + r-1)} p_{m+1} , \label{split3}
\end{eqnarray}
and we see that $\xi$ is of order $O(1/K)$. Therefore, Eq.(\ref{split1}) becomes 
\begin{eqnarray}
\dot p_m &=& \frac{b (m+1) (r + m + \rho )}{m+1  + r} p_{m+1} -  \frac{ b m  (r + m-1 + \rho )}{m  + r} p_{m} -  \frac{ a (m + r \epsilon)  }{m  + r} p_{m}  +  \frac{a (m-1 +\epsilon r )  }{m-1  + r} p_{m-1} \nonumber\\
&&+  O(1/K), \label{bmaster4}
\end{eqnarray}
which is Eq.(\ref{master4}) in the limit of $K \to \infty$.
More rigorously, the coupled equations Eqs.(\ref{split1}) and (\ref{split2}) reduce to one equation in the limit of $K \to \infty$, due to  Tikhonov's theorem on dynamical system~\cite{tikh,klon}, where  $\xi$ is obtained from Eq.(\ref{split3}) after setting $K^{-1}$ to zero,  and then substituted into Eq.(\ref{split1})  to obtain Eq.(\ref{master4}). 

\section{The leading order contributions to $|\lambda_1|$ and $v^{(k \ge 2)}_0$ when $\lambda_1/\lambda_{k \ge 2} \gg 1$}\label{exp}
Let us consider the transition rate matrix ${\bf K}$ for Eq.(\ref{master1}) or Eq.(\ref{master4}) with $\epsilon=0$, whose $(i,j)$-th element is $k_{i \to j}$, so that the probability distribution is represented as a row vector, and the time derivative is obtained by multiplication of the transition matrix from the right. For the current model, the transition matrix takes the form
\begin{equation}
%\[
\raisebox{-5pt}{{\huge\mbox{{${\bf K}$}}}} = \left(
\begin{array}{c|c}
  0 & 0 \ \ \cdots \ \ 0 \\ \hline
  \alpha & \raisebox{-15pt}{{\huge\mbox{{${\bf A}$}}}} \\[-3ex]
  0 & \\ [-2ex]
  \vdots & \\[-2ex]
  0 &
\end{array}
\right)\label{tmat}
%\] 
\end{equation}
where the first state is taken to be the absorbing state, whose index is taken to be zero, and ${\bf A}$ is the submatrix formed by the transition rates between the other states, whose indices are $m=1,2, \cdots$. For Eq.(\ref{master4}), $\alpha \equiv k_{1 \to 0} = b (r + \rho) / (1+r) $. 

We see that $\langle {{\bf v}^{(0)}} |=(1,0,0 \cdots )$ is the left eigenvector of ${\bf K}$ with the eigenvalue 0, the stationary state. Because $\sum_j k_{i \to j} = 0$, we have $ {\bf K} | {\bf I} \rangle = {\bf 0}$ where $| {\bf I} \rangle \equiv (1,1, \cdots 1)^T$. This also tells us that for any left eigenvector  $\langle {\bf v} | = (v_0, v_1, \cdots)$  for a non-zero eigenvalue $\lambda$, we have
\begin{equation}
\lambda \langle {\bf v} | {\bf I} \rangle =  \langle {\bf v} | {\bf K} | {\bf I} \rangle = 0,
\end{equation}
leading to 
\begin{equation}
\sum_i v_i = 0. \label{sumzero}
\end{equation}
Also, because of the special form of $\bf K$ given in Eq.(\ref{tmat}), expressing the left eigenvector as $\langle {\bf v} | = v_0 \oplus \langle {\bf \tilde v} | $ where $\langle {\bf \tilde v} | = (v_1, v_2 \cdots )$, we have 
\begin{equation}
\langle {\bf v} | {\bf K} = \alpha v_1 \oplus \langle {\bf \tilde v} | {\bf A} = \lambda v_0 \oplus \lambda \langle {\bf \tilde v} |,
\end{equation}
which shows  that $\lambda$ is also an eigenvalue of ${\bf A}$ with the corresponding left eigenvector $\langle {\bf \tilde v} |$, and 
\begin{equation}
\lambda = \alpha \frac{v_1}{v_0} = -  \frac{\alpha v_1}{\sum_{m \ge 1} v_m}. \label{egvn}
\end{equation}

 When we set $\alpha$ to zero, the corresponding modified transfer matrix ${\bf K}_0$ describes the Markov model in Eq.(\ref{mrm}), where $\bf A$ is replaced by ${\bf A}_0$, defined as
\begin{equation}
{\bf A}_0 = {\bf A} + \alpha {\bf P},\label{proj}
\end{equation}
where ${\bf P}$ is a projection matrix with the definition $P_{i j} \equiv \delta_{i 1} \delta_{j 1}$. The submatrix ${\bf A}_0$ possesses the left eigenvector $\langle {\bf p}^{\rm qs}|$ with the zero eigenvalue, satisfying
 \begin{equation}
 \langle {\bf p}^{\rm qs} | {\bf A}_0 = 0, \label{zero}
 \end{equation}
which we called the quasi-steady distribution in the main text. The eigenvalues and eigenvectors of ${\bf K}$ can be obtained from those ${\bf K}_0$ of by the perturbations of size $O(\alpha/A)$ where $A$ is the typical size of ${A_0}_{ij}$ that determines the sizes of the nonnegative eigenvalues of ${\bf A}_0$. In particular, the left eigenvector $\langle {\tilde {\bf v}}^{(1)} |$ of $\bf A$ for the eigenvalue $\lambda_1$ is obtained from $\langle {\bf p}^{\rm qs}|$ as
\begin{equation}
\langle {\tilde {\bf v}}^{(1)} | =  \langle {\bf p}^{\rm qs}| + O(\alpha A^{-1}). \label{pert}
\end{equation}
From Eqs.(\ref{egvn}) and (\ref{pert}), we have
\begin{equation}
-\lambda_1 =   \frac{\alpha v^{(1)}_1}{\sum_{m \ge 1} v^{(1)}_m} = \frac{\alpha  p^{\rm qs}_1}{\sum_m p^{\rm qs}_m } + O(\alpha A^{-1}),
\end{equation}
where we see that the first term in the final expression is nothing but $\tau_q^{-1}$ given in Eq.(\ref{leakrate}). The corresponding eigenvector in the full state space is 
\begin{equation}
\langle {\bf v}^{(1)} | =  (-\sum_{m \ge 1} p^{\rm qs}_m) \oplus\langle {\bf p}^{\rm qs}|  + O(\alpha A^{-1}),\label{egv}
\end{equation}
where Eq.(\ref{sumzero}) was invoked.  

  The eigenvectors for $\lambda_{k \ge 2}$ are obtained from those for the negative eigenvalues of ${\bf A}_0$. Because ${\bf A}_0$ is a transition rate matrix in the subspace of $m \ge 1$ states, a left eigenvector $\langle \tilde {\bf v} | = (v_1, \cdots) $  of ${\bf A}_0$ for an eigenvalue $\lambda < 0$ satisfies the equation $\sum_{i \ge 1} v_i = 0$. Therefore, we see that for an eigenvector $\langle {\bf v}^{(k)} | = (v^{(k)}_0, v^{(k)}_1, \cdots )^T$ for the eigenvalue $\lambda_k$ with $k \ge 2$, we have
\begin{equation}
\sum_{m \ge 1} v^{(k)}_m \sim O(\alpha A^{-1}).
\end{equation}
Consequently,
\begin{equation}
\lambda_k = - \frac{ \alpha v^{(k)}_1}{\sum_{m \ge 1} v^{(k)}_m} \sim O(A), 
\end{equation} 
and 
\begin{equation}
v^{(k)}_0 = -\sum_{i \ge 1} v^{(k)}_m \sim O(\alpha A^{-1}). \label{zerocomp}
\end{equation}
for $k \ge 2$. From Eq.(\ref{zerocomp}), we see that $v^{(k)}_0$ for $k \ge 2$ can be neglected if $\alpha/A$ is small enough. In this case, if  we start from the initial condition with $p_0(0)=0$, we have $p_0(0)=1 + \sum_{k \ge 1} v^{(k)}_0  \simeq  1 + v^{(1)}_0 = 0$, leading to $v^{(1)}_0 \simeq -1$, and consequently the leading order contribution to the stationary state is of the form $p_0(t) \simeq 1 - \exp(-t/\tau_q)$ with no further dependence on the initial condition.

\section{The asymptotic form of $p_0(t)$ for $K/b=0$ and $\epsilon=0$}\label{free}
The bound and free modes are completely decoupled for $K/b=0$, and their probabilities are  conserved separately. Restricting to the free mode, the eigenvalues for the approach to the stationary distribution can be obtained in an analytic form when $\epsilon=0$. The transition rate matrix for the free mode is of the form
\[
K=\begin{bmatrix}
0 & 0 & 0 & 0 & \cdots  \\
b & -b & 0 & 0 & \cdots \\
0 & 2b & -2b & 0 & \cdots \\ 
0 & 0 & \ddots & \ddots & \ddots
\end{bmatrix}. 
\]
 From the form of the matrix, it is easy see that the eigenvalues are 0, $-b$, $-2b$, $-3b$, $\cdots$. Similarly, from the matrix for the bound mode, where $b$ in the expression above is replaced by $\rho b$, we see that the corresponding eigenvalues are 0, $-\rho b$, $-2\rho b$, $-3\rho b$, $\cdots$. For both the free and the bound modes, we see that sizes of the eigenvalues are not well separated. In other words, the time-scale separation does not hold, and $1-p_0(t)$ exhibits initial condition dependent multi-exponential behavior,
\begin{equation}
1-p_0(t) = \sum_k A_k e^{-k b t} + \sum_k B_k e^{-k \rho b t},
\end{equation}
where the constants $A_k$s and $B_k$s are determined by the initial condition. Even if we assume $p_0(0)=0$, these constants are not fully determined. For the special case of $\rho=0$, we have the form
\begin{equation}
1-p_0(t) = \sum_k A_k e^{-k b t} +  B,
\end{equation}
and 
$1-p_0(t) \simeq A_1 e^{-b t} +  B$ for  $bt \gg 1$. Note that $1-p_0(t)$ is approximated by a single-exponential form only for $bt \gg 1$ where $1-p_0(t) \ll 1$, because the time-scales are not well separated.
\section{Derivation of Fokker-Planck equation from the master equation.}\label{FP}
The chemical master equation for single species can be written as~\cite{elf,kampen2,dyken}
\begin{equation}
\partial_t  P(n,t) = \bar m \sum_{j=1}^R (F_j (n-S_j) -  F_j(n)), \label{master5}
\end{equation}
where $j=1, \cdots R$ labels reactions in the system.  In Eq. (\ref{master5}), $F_j(n)$ and $S_j$ are the transition rate and  the increase of the particle number, respectively, for the $j$-th reaction, and $\bar m$ is the size parameter, a large number whose size is comparable to the average protein number. For the reduced master equation Eq.(\ref{master4}), we have two reactions, the creation and the degradation, with $S_1=1$ and $S_2=-1$, and
\begin{eqnarray}
F_1(m) &=& \frac{a (m+r \epsilon)}{\bar m (m+r)}  \nonumber\\
F_2(m) &=&  \frac{b m (m+r + \rho -1 )}{\bar m (m+r)}.  \label{trnftn}
\end{eqnarray}

When the average number of protein molecules is large, one can approximate the discrete variable $x \equiv m/\bar m$ as a continuous variable. Considering $F_j$ as functions of $x$, $f_j(x) \equiv F_j(n)$,  with $\pi(x,t) \equiv \bar m p_m(t)$, we get the Kramer-Moyal expansion~\cite{kampen2}
\begin{eqnarray}
\partial_t  \pi(x,t) &=& \bar m \sum_{j=1}^R \left(f_j (x-\frac{S_j}{\bar m}) \pi (x-\frac{S_j}{\bar m},t)  -  f_j(x)\pi (x,t)\right)\nonumber\\
 &=& \sum_{j=1}^R \sum_{k=1}^\infty \bar m^{1-k}(-S_j)^k \partial_x^k (f_j(x) \pi (x,t))\nonumber\\
 &=&   \sum_{k=1}^\infty (-1)^k \bar m^{1-k} \partial_x^k  (a_k(x) \pi (x,t)).
\label{expan1}
\end{eqnarray}
where $a_k(x) \equiv \sum_j S_j^k f_j(x)$.
Assuming that $\bar m$ is large enough so that the expansion can be kept only up to the second order, we obtain the Fokker-Planck equation~\cite{kepler,kampen2}
\begin{equation}
\dot \pi(x,t) = -\partial_x(A(x) \pi(x,t)) + \frac{1}{2}\partial_x^2 (B(x) \pi(x,t))\label{FP2}.
\end{equation}
where $A(x) \equiv a_1(x)$ and $B(x) \equiv \bar m^{-1} a_2(x)$. For Eq.(\ref{trnftn}), we get
\begin{eqnarray}
f_1(x) &=& \frac{\tilde a (x + \tilde r \epsilon)}{x + \tilde r} \nonumber\\
f_2(x) &=& \frac{b x (x + \tilde r + \bar m^{-1} (\rho -1))}{x + \tilde r} \nonumber\\
&=& b x + \frac{b x (\rho -1)}{\bar m (x + \tilde r)} 
\end{eqnarray}
with the rescaled rates $\tilde r = r /\bar m$ and $\tilde a = a /\bar m$. 
Therefore, we have
\begin{eqnarray}
A(x) &=&  \frac{\tilde a (x + \tilde r \epsilon)}{x+\tilde r} -b x  +\frac{b x (1-\rho)}{\bar m (x+\tilde r)}\nonumber\\  
B(x) &=& \frac{1}{\bar m}  \left( \frac{\tilde a (x + \tilde r \epsilon)}{x+ \tilde r}+  b x \right)+ O(\bar m^{-2}).  \label{FP3}
\end{eqnarray}

Note that, in the limit of $\bar m \to \infty$, Eq.(\ref{FP2}) reduces to
 \begin{equation}
 \partial_t \pi(x,t) = -\partial_x\left[ \left(-b x + \frac{\tilde a (x + \epsilon \tilde r) }{x+\tilde r}\right) \pi(x,t) \right]. \label{de1}
 \end{equation}
Because there is no diffusion term in Eq.(\ref{de1}), uncertainty originates purely from the initial condition. Therefore, the dynamics described by Eq.(\ref{de1}) is deterministic, and is equivalent to the  rate equation
\begin{equation}
\dot x = \frac{\tilde a (x + \epsilon \tilde r)}{x+\tilde r} - b x. \label{dg}
\end{equation}

Note that $\bar m$ is chosen to be of size comparable to the average number of proteins, implying that it is reasonable to take $\bar m$ to be $O(m^*)$ where $m^*$ is the position of the nonzero peak of the quasi-steady distribution. Because $m^* \simeq a/b - r$ when $a/b-r \gg 1$, we may simply define $\bar m \equiv a/b-r$. With this definition of $\bar m$, we have $\tilde a = ab/(a-rb )$, $\tilde r = rb/(a-rb)$, and the stable fixed point of the rate equation is $x_1=1$. 

The parameter $\rho$ does not appear in the Eq.(\ref{dg}) because the average number of the protein molecules is much larger than unity and therefore most of the protein molecules are in the free form. Consequently, the degradation of the bound protein molecule gives negligible contribution to the overall degradation of proteins in the limit where the deterministic rate equation is valid. Similarly, by fixing $\tilde r =  k_1/(k_0 \bar m)$ to a finite value,  the unbinding rate $k_1$ in the chemical master equation is taken to be much larger than the binding rate $k_0$  when $\bar m \gg 1$. Therefore, the inactivation of the DNA due to the degradation of the bound protein is negligible compared to that due to the unbinding.
% This can be understood from the fact that for finite values of $\tilde a$ and $\tilde r$, $a=\bar m \tilde a \to \infty$ and $r = \bar r \tilde m \to \infty$ as $\bar m \to \infty$.  

\section{The shift of the stable fixed point in the stationary distribution and the stochastic slow down}\label{shiftx}
It has been noted that when the production rate is a function of protein numbers that is concave downward, than the production rate slows down due to stochastic noise, which in turn decreases the average value of the protein number of the steady state relative to the value obtained by the deterministic rate equation~\cite{dyken}. We sketch the derivation for the shift of the steady state average number of the proteins below. Although the multi-species master equation was considered in ref.~\cite{dyken}, we restrict ourselves to the single species case for notional simplicity. The shift of the average value was obtained by expanding Eq.(\ref{master5}) around the solution $\bar x$ that satisfies the deterministic rate equation,
\begin{equation}
\frac{d \bar x}{d t} = \sum_j S_j f_j(\bar x), \label{dagn}
\end{equation}
 and considering the probability density $\Pi (\epsilon, t) = \bar m^{1/2} P(n,t)$~\cite{kampen2,dyken}~\footnote{The factor of $\bar m^{1/2}$ is absent in Eq.(9) of Ref.\cite{dyken}, but it is required for relating the probability mass function of a discrete variable to the probability density of a continuous variable. This factor cancels out in the left and the right-hand of the master equation, so the master equation remains unchanged.}. After substituting 
\begin{equation}
x = \bar x + {\bar m}^{-1/2} \epsilon. 
\end{equation}
into Eq.(\ref{master5}), the expansion to order $\bar m^{-1/2}$ results in the equation~\cite{dyken}
%\footnote{The sign of the $J_{ik}$ term in Eqs.(12),(13),(14), and (15) of ref.\cite{dyken}) are flipped, which is corrected here.}
\begin{eqnarray}
\partial_t \Pi (\epsilon, t) &=& - \sum_j S_j f_j'(\bar x) \partial_\epsilon (\epsilon \Pi (\epsilon, t)) + \frac{1}{2} \sum_j S_j^2 f_j (\bar x) \partial_\epsilon^2 \Pi (\epsilon, t) \nonumber\\
&&+ \bar m^{-1/2} (- \sum_j S_j f_j''(\bar x)\partial_\epsilon(\epsilon^2 \Pi (\epsilon, t)) + \sum_j S_j^2 f_j' (\bar x) \partial_\epsilon^2(\epsilon \Pi (\epsilon, t)), \label{sizeexp}
\end{eqnarray}

By multiplying the both sides of Eq.(\ref{sizeexp}) by $\epsilon$ and integrating over $\epsilon$, one obtains~\cite{dyken}
\begin{equation}
\langle \dot  \epsilon \rangle = \sum_j S_j f_j'(\bar x) \langle \epsilon \rangle + \bar m^{-1/2} \frac{1}{2} \sum_j S_j f_j''(\bar x) \langle \epsilon^2 \rangle  + O(\bar m^{-1}).
\end{equation}
where the brackets denote the average value, and the term of $O(\bar m^{1/2})$ was removed by using the equation Eq.(\ref{dagn}). 
Therefore, for stationary state where $ \langle \dot \epsilon \rangle = 0$, we get the leading order shift
\begin{equation}
\langle x \rangle = \bar x - \bar m^{-1} \frac{\sum_j S_j f_j''(\bar x) \langle \epsilon^2 \rangle}{2 \sum_j S_j f_j'(\bar x)}.
\end{equation}
When $\bar x$ is the stable fixed point, then $\sum f_j'(\bar x) < 0$, and the shift is negative if $f_n''(\bar x)$, as in the case of Michelis-Menten type production rate and linear degradation rate~\cite{dyken}, which is also the case for our model. That is, the average value of the particle number of the stationary distribution is less than the stable fixed point, to the leading order in stochastic noise. 

We note one subtle point. In ref.\cite{dyken}, $\bar m$ dependence of $f_j(n)$ was not considered. In our model, $f_2(n)$ in fact contains a term of $O(\bar m^{-1})$ unless $\rho=1$. Therefore, Eq.(\ref{dagn}) is not exactly the same as the deterministic rate equation we considered previously, where the $\bar m^{-1}$ dependent term was dropped. Therefore, we now have to consider the stable fixed point $\bar x$ of Eq.(\ref{dagn}) that can be written as
\begin{equation}
\dot x =  \frac{\tilde a (x + \epsilon \tilde r)}{x+\tilde r} -b x +  \frac{ b x (1-\rho) }{\bar m ( x+\tilde r)} \label{dmixed}
\end{equation}
for our model, and examine the {\it additional} position shifted due to the stochastic noise. Let us also consider $\epsilon=0$ and examine the effect of the shift both due to the stochastic noise and the baseline production. For  $\epsilon=0$ ,  Eq.(\ref{dmixed}) is then written as
\begin{equation}
\dot x =  \frac{\bar a x }{x+\tilde r} -b x  \label{dmixed2}
\end{equation}
with $\bar a \equiv \tilde a + b ( 1- \rho) \bar m^{-1}$, so it takes the same form as the deterministic rate equation considered previously, with redefinitions of the parameters. Therefore, the non-zero fixed point of Eq.(\ref{dmixed2}) is $\bar x = \bar a/b - \tilde r$. 

We already considered the shift of the unstable fixed point in the main text, in the context of the Fokker-Planck equation. The equation Eq.(\ref{fd2}) was expanded around a fixed point of Eq.(\ref{de2}) to obtain Eq.(\ref{fd3}) that describes the shift of an extremum. If we instead expand Eq.(\ref{fd2}) around the fixed point of Eq.(\ref{dmixed2}), we obtain
\begin{equation}
(-2 b \bar x -b \tilde r  + \bar a) \delta x = -a \tilde r \epsilon  + \frac{1}{2 \bar m } \left( \frac{ \tilde a \tilde r }{\bar x + \tilde r } + b (\bar x+\tilde r) \right) +O({\bar m}^{-2},\epsilon^2,{\bar m}^{-1} \epsilon). \label{fd4}
\end{equation}
and therefore
\begin{eqnarray}
\delta x &=& (-2 b \bar x -b \tilde r  + \bar a)^{-1} \left( -a \tilde r \epsilon  + \frac{1}{2 \bar m } \left( \frac{ \tilde a \tilde r }{\bar x + \tilde r } + b (\bar x+\tilde r) \right)  \right)\nonumber\\
&=& (b \tilde r  - \bar a)^{-1} \left( -a \tilde r \epsilon  + \frac{1}{2 \bar m } \left( \frac{ \tilde a b \tilde r }{\bar a } + \bar a \right)  \right).
\end{eqnarray}
Therefore, again, we see that the effect of the stochastic noise and the baseline production is opposite. The former tends to shift the  maximum to negative direction whereas the baseline production tends to shift it in positive direction. When the stochastic noise is small and the non-zero peak is dominant, its position is approximately the average particle number. When the peak at the zero gives sizeable contribution to the probability distribution, then the average particle number is less than the position of the non-zero peak. Therefore, the negative shift of the peak relative to the stable fixed point of the Eq.(\ref{dmixed}) is consistent with the result of ref.\cite{dyken}, which states that the average number of particles are less than the stable fixed point of Eq.(\ref{dmixed}). 
\newpage

\end{document}